\pdfoutput=1
\documentclass[12pt]{iopart}
\usepackage{bm}
\usepackage{graphicx}
\usepackage{color}
\usepackage{subfigure}
\usepackage{hyperref}
\usepackage{latexsym}
\usepackage{amsthm}
\usepackage{amssymb}
\usepackage{cancel}

\usepackage{cite}

\DeclareGraphicsExtensions{.jpg,.pdf, .mps, .png, .eps, .ps, .EPS,.gif}

\begin{document}
\newcommand{\updownarrows}{\mathbin\uparrow\hspace{+.0em}\downarrow}
\newcommand{\downuparrows}{\mathbin\downarrow\hspace{-.5em}\uparrow}
\def\be{\begin{equation}}
\def\ee{\end{equation}}

\def\bc{\begin{center}}
\def\ec{\end{center}}
\def\bea{\begin{eqnarray}}
\def\eea{\end{eqnarray}}
\newcommand{\avg}[1]{\langle{#1}\rangle}
\newcommand{\Avg}[1]{\left\langle{#1}\right\rangle}

\def\ie{\textit{i.e.}}
\def\etal{\textit{et al.}}
\def\m{\vec{m}}
\def\G{\mathcal{G}}

\title[The  higher-order spectrum of  simplicial complexes: a RG approach]{ The  higher-order spectrum of  simplicial complexes: \\ \ a renormalization group approach}

\author{Marcus Reitz}

\address{Institute for Mathematics, Astrophysics and Particle Physics, Radboud Universiteit, Heyendaalseweg 135, NL-6525 AJ Nijmegen, The Netherlands}
\ead{m.reitz@science.ru.nl}
\author{Ginestra Bianconi}
\address{School of Mathematical Sciences, Queen Mary University of London, London, E1 4NS, United Kingdom\\
Alan Turing Institute, the British Library, London, United Kingdom}
\ead{g.bianconi@qmul.ac.uk}
\vspace{10pt}
\begin{indented}
\item[]
\end{indented}

\begin{abstract}
Network topology is a flourishing interdisciplinary subject that is relevant for different disciplines including  quantum gravity and brain research. The discrete topological objects that are investigated in network topology are simplicial complexes. Simplicial complexes generalize networks by not only taking pairwise interactions into account, but also taking into account many-body interactions between more than two nodes.
Higher-order Laplacians are topological operators that describe higher-order diffusion on simplicial complexes and constitute the natural mathematical objects that capture the interplay between network topology and dynamics. We show that higher-order up and down Laplacians can have a finite spectral  dimension, characterizing the long time behaviour of the diffusion process on  simplicial complexes that depends on their order $m$. We  provide a renormalization group theory for the calculation of the higher-order spectral dimension of two  deterministic models of simplicial complexes: the  Apollonian and the pseudo-fractal simplicial complexes. We show that the RG flow is affected by the fixed point at zero mass, which determines the higher-order spectral dimension $d_S$ of the up-Laplacians of order $m$ with  $m\geq 0$. 
\end{abstract}

%
%
%
%
%

\section{Introduction}
Simplicial complexes \cite{Perspective,Bassett,Lambiotte1,top1,Equilibrium,Costa} are generalized network structures that capture many-body interactions. They are not just formed by nodes and links like networks but they also include  simplices of higher dimensions such as  triangles, tetrahedra and so on.
Being build by these topological building blocks, simplicial complexes are the  ideal discrete structures to investigate emergent geometry \cite{Emergent, Hyperbolic, Flavor, Polytopes,Bath} and can be described by discrete algebraic and combinatorial topology. Topology  is a traditional tool of high-energy physics and quantum gravity and recently it has also become increasingly popular  to investigate complex systems\cite{Chris}. In fact  topological methods have been shown to be very powerful to analyse datasets, including brain networks and  collaboration networks \cite{Vaccarino,Aste1,Bassett,Nanoparticles}. Finally there is an increasing interest in revealing the role that the higher-order interactions of simplicial complexes have on their dynamics \cite{Barbarossa,High_Laplacian,Arenas,Iacopini,Kahng_SIS,Arenas2}.

The network Laplacian  \cite{Chung,Samukhin1,Samukhin2,Corona_chinese} is fundamental to understand the interplay between topology and dynamics and its spectral properties are known to affect diffusion and synchronization on network structures.
In particular the spectral dimension \cite{Toulouse,RG1,Kahng,Kim,Burioni1,Burioni2,Burioni3,Jonsson1,Jonsson2,Nice} characterizes the spectral properties of networks with distinct geometrical features and determines the late time behavior of diffusion and more general dynamical processes on networks \cite{Ising,Erzan,Ana,Ana2,Bialek}.
The spectral dimension can also be defined on simplicial complexes \cite{RG1} by focusing on their skeleton (the network obtained from a simplicial complex by retaining only its nodes and links).  Thus, the  spectral dimension is also considered a key mathematical object for investigating the effective dimension of a simplicial quantum geometry as felt by diffusion processes. More in general in quantum gravity the spectral dimension is used  for probing the geometry of the simplicial spacetimes \cite{Loll2,dario1,dario2} described by different theoretical approaches including Causal-Dynamical-Triangulations (CDT) \cite{Loll}.

Here we focus on two models of pure $d$-dimensional simplicial complexes called  Apollonian simplicial complexes,\cite{apollonian,apollonian2,apollonian_group} and pseudo-fractal simplicial complexes \cite{pseudo_Doro}.
The Apollonian simplicial complexes \cite{apollonian,apollonian2,apollonian_group} are  deterministic hyperbolic $d$-dimensional manifolds that are obtained by an iterative process, whose limit converges to an infinite hyperbolic lattice. The Apollonian simplicial complex in $d$-dimensions is dominated by the boundary and is  closely related to the melonic graphs of tensor networks  \cite{Tensor1,Tensor2}, because melonic graphs can be understood as the merging of two identical Apollonian simplicial complexs upon identification of the all their faces at the boundary.
The pseudo-fractal simplicial complexes \cite{pseudo_Doro}, generalise the Apollonian simplicial complexes to simplicial complexes that are not manifolds. These deterministic simplicial complexes  have a skeleton which is non-amenable, i.e. they have an infinite isoperimetric dimension and simultaneously have a very small Cheeger constant \cite{Cheeger1,Rosenthal}. Additionally, they are small world and scale-free.

 {While the Apollonian and the pseudo-fractal simplicial complexes are generated iteratively by a deterministic algorithm, most of the real networks are the outcome of a stochastic process.
In is therefore important to note that the two classes of simplicial complexes considered here constitute the backbone of the more general simplicial complex model called  ``Network Geometry with Flavor"  \cite{Hyperbolic,Flavor,Bath}.
This model generates random simplicial complexes whose structure evolves according to a stochastic process, where the set of possible simplices is restricted to be a depending on the model parameters either a subset of the faces of the Apollonian simplicial complex  or a subset of the pseudo-fractal simplicial complexes. }

Given the fact that Apollonian and pseudo-fractal simplicial complexes are highly geometrical, deterministic and hierarchical, these structures  and their generalizations \cite{Havlin1,Havlin2} are very suitable for conducting renormalization group (RG) calculations analytically. Examples of dynamical  processes already studied with the RG in related simplicial complex models  include percolation\cite{BoettcherZiff,Patchy,percolation_apollonian,BianconiZiff1,BianconiZiff2,BianconiZiff3}, spin models \cite{Boettcher_RG} and Gaussian models \cite{RG1,Kahng,Kim}.

In this paper  we investigate the properties of higher-order Laplacians \cite{simplices2,thesis,Jost,High_Laplacian,Cheeger1,Rosenthal} on the considered simplicial complexes.
The higher-order Laplacians describe diffusion processes occurring on higher-order simplices \cite{simplices2,High_Laplacian} and are key mathematical objects to define the higher-order Kuramoto model \cite{Explosive}. Higher order Laplacians are also closely related to approximate Killing vector fields, which are currently being investigated on quantum geometries in CDT \cite{Killing}. 
It has been recently shown numerically \cite{High_Laplacian}, that the higher-order up-Laplacian and down-Laplacian can display  a finite spectral dimension. 
Here we use renormalization group (RG) theory  \cite{RG1,Kahng,Kim} to analytically calculate the spectral dimension of higher-order up-Laplacians of Apollonian and pseudo-fractal simplicial complexes. 
We find that each simplicial complex belonging to the considered class of models, is characterized by a set of analytically predicted spectral dimensions. Each spectral dimension corresponds to the spectrum of a  higher-order up-Laplacian of different order $m$. The values of the predicted spectral dimensions are compared to direct numerical results for $d=3$ and $d=4$ simplicial complexes. 

The paper is structured as follows: in Sec. II we  introduce simplicial complexes and their higher-order Laplacians, in Sec.III we  present the hyperbolic and non-amenable simplicial complex models considered in this work; in Sec. III we  give the necessary background for deriving the higher-order spectrum of the Apollonian and pseudo-fractal simplicial complexes using the RG approach; In Sec. IV and in Sec. V we derive the RG equations and the RG flow for the Apollonian simplicial complexes;  In Sec. VI and Sec. VII we derive the RG equations and the RG flow for the pseudo-fractal  simplicial complexes. In Sec. VIII we summarize the main analytical results and we will compare with numerical results on all the considered simplicial complex models.
Finally, in Sec. IX we will provide the conclusions.

\section{Simplicial complexes and higher-order Laplacians }

\subsection{Simplicial complexes}
A {\em $m$-dimensional simplex} $r$ (also indicated as $m$-simplex)  includes $m+1$ nodes and it can be indicated as
\bea
r=[v_0,v_1,\ldots, v_{m}].
\eea
Therefore, a $0$-simplex is a node, a $1$-simplex is a link, a $2$-simplex a triangle, a $3$-simplex a tetrahedron and so on.
A $m^{\prime}$-dimensional face $q$ of a $m$-dimensional simplicial complex $r$ is a $m^{\prime}<m$ simplex formed by a subset of $m^{\prime}+1$ nodes belonging to the simplex $r$.

In topology, simplices also have an {\em orientation}. Two $m$-simplices differing only by the order in which their nodes are listed are therefore related by 
\bea
[v_0,v_1,\ldots, v_{m}]=(-1)^{\sigma(\pi)}[v_{\pi(0)},v_{\pi(1)},\ldots, v_{\pi(m})],
\eea
where $\sigma(\pi)$ indicates the parity of the permutation $\pi$ of the $m+1$ indices of the nodes.

A {\em simplicial complex} is formed by a set of simplices  with the property that the simplicial complex is closed under inclusion of the faces of any of its simplices.

A  {\em $d$-dimensional simplicial complex} is a simplicial complex for which the maximum dimension of its simplices is $d$. Here we are exclusively interested in {\em pure $d$-dimensional simplicial complexes}, which are formed by a set of $d$-dimensional simplices and all their faces.
The {\em skeleton} of a simplicial complex is the network formed by the set of all the nodes and links of the simplicial complex.
Given a $d$-dimensional simplex,  we indicate  the number of its  $m$-simplices with $N^{[m]}$ with $0\leq m\leq d$.

\subsection{Boundary map and incidence matrices}
Given a simplicial complex, a {\em $m$-chain} consists of the elements of a free abelian group $\mathcal{C}_{m}$ {with basis  formed by the set of all $m$-simplices of the simplicial complex.
Therefore every element $a\in \mathcal{C}_m$ can be  uniquely expressed as a linear combination of basis elements  with  coefficients given $c_r\in \mathbb{Z}_2$, i.e.  }
\bea
a=\sum_{r\in {Q}^{[m]}}c_r \left[v_0^{(r)},v_1^{(r)},\ldots, v_{m}^{(r)}\right]
\eea 
 {where $c_r\in \{1,-1\}$. Here  ${Q}^{[m]}$ indicates the set of all $m$-simplices of the simplicial complex and each $m$-simplex $r\in {Q}^{[m]}$ of the simplicial complex  is indicated by  $r=[v_0^{(r)},v_1^{(r)},\ldots, v_{m}^{(r)}]$.}

The {\em boundary map} $\partial_{m}$ is a linear operator $\partial_{m}:\mathcal{C}_{m}\to \mathcal{C}_{m-1}$ whose action is determined by the action on each $m$-simplex of the simplicial complex.
In particular the boundary map $\partial_{m}$ applied to the $m$ simplex $r=[v_0,v_1,\ldots, v_{m}]$ gives 
\bea
\partial_{m}[v_0,v_1\ldots,v_{m}]=\sum_{j=0}^{m}(-1)^j[v_0,v_1,\ldots, v_{j-1},v_{j+1},\ldots, v_{m}].
\label{boundary}
\eea
In words, the boundary map applied to a $m$-simplex gives a linear combinations of its $(m-1)$-dimensional faces.

We say that two $m$-faces $r $ and $q$ of a simplicial complex are upper adjacent  if there is a  $(m+1)$-simplex $\tau$ of which both $r$ and $q$ are faces.
The $m$-faces $r$ and $q$ are upper adjacent with similar  orientation if the simplicial complex contains a $(m+1)$-dimensional simplex $\tau
$ such that \bea
\langle{r,\partial_{m+1}\tau}\rangle=\langle{q,\partial_{m+1}\tau}\rangle,
\eea 
where $\langle a,b\rangle$  indicates the inner product  {on $\mathcal{C}_{m}$}.
Conversely, they are upper adjacent with opposite orientation if the simplicial complex contains a $(m+1)$-dimensional simplex $\tau
$ such that \bea
\langle{r,\partial_{m+1}\tau}\rangle=-\langle{q,\partial_{m+1}\tau}\rangle.
\eea 

From the definition of the boundary map $\partial_{m}$ given by Eq. (\ref{boundary}), it follows immediately that for every $m$-dimensional simplex $r$
\bea
\partial_{m-1}\partial_{m}r=0,
\label{boundary2}
\eea
which is an important topological property that can be expressed in words with the sentence ``the boundary of a boundary is null".

\begin{figure}[h!]
	\begin{center}
 \includegraphics[width=0.7\columnwidth]{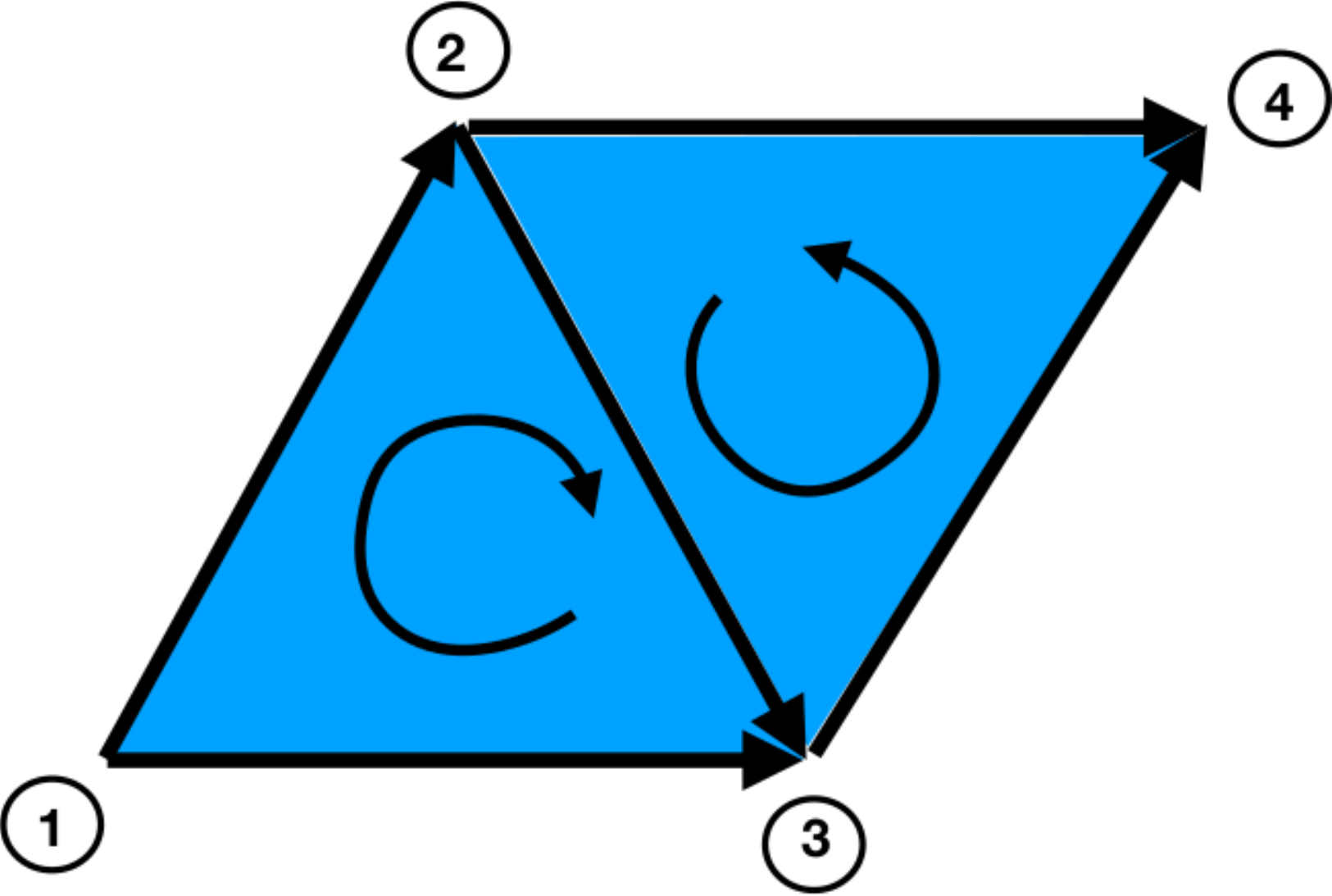}	
  \end{center}\caption{ An example of a small simplicial complex with the orientation of the simplices induced by the labelling of the nodes.}
  	\label{fig:example}
	\end{figure}

Given a simplicial complex with $N^{[m]}$ $m$-dimensional simplices we can  choose a  base for  ${\mathcal C}_{m}$  by taking  an  ordered list of its $m$ simplices.
If we fix both the base of ${\mathcal C}_{m}$ and ${\mathcal C}_{m-1}$ we can represent the boundary operator $\partial_m$ by a $N^{[m-1]}\times N^{[m]}$ {\em incidence matrix} ${\bf B}_{[m]}$.
In Figure $\ref{fig:example}$ we show an example  of a simplicial complex. We choose as  bases for ${\mathcal C}_{0},{\mathcal C}_{1}$ and ${\mathcal C}_{2}$ the ordered list of nodes $\{[1],[2],[3],[4]\}$, links  $\{[1,2],[1,3],[2,3],[3,4],[2,4]\}$ and triangles $\{[123],[234]\}$.
With this choice of bases, the boundary maps $\partial_1$ and $\partial_2$  can be represented by the incidence matrices ${\bf B}_{[1]}$ and ${\bf B}_{[2]}$  with,
\bea
{\bf B}_{[1]}=\left(\begin{array}{ccccc}
-1&-1 &0&0&0\\
1&0&-1&0&-1\\
0&1&1&-1&0\\
0&0&0&1&1\\
\end{array}\right),
\ 
{\bf B}_{[2]}=\left(\begin{array}{cc}
1&0\\
-1& {0}\\
1& {1}\\
0&1\\
 {0}& {-1}

\end{array}\right).
\eea

\subsection{Higher order Laplacian matrices of simplicial complexes}

The graph Laplacian or $0$-Laplacian describes the diffusion process over a network and it is an extensively studied topological operator in graph theory \cite{Chung}.
The $0$-Laplacian can be also defined for a simplicial complex and describes the diffusion process that goes from a node to another node across shared links.
In fact the  $0$-Laplacian is a $N^{[0]}\times N^{[0]}$ matrix and can be expressed in terms of the incidence matrix ${\bf B}_{[1]}$, 
\bea
{\bf L}_{[0]}={\bf B}_{[1]}{\bf B}^{\top}_{[1]}.
\eea
While on networks only the graph Laplacian and its normalized versions can be defined, in simplicial complexes it is possible to define higher-order Laplacians describing diffusion taking place between higher-order simplices.
The higher-order Laplacian ${\bf L}_{[m]}$ with $m>0$ (also called combinatorial Laplacians)  can be represented as a $N^{[m]}\times N^{[m]}$ matrix given by 
\bea
{\bf L}_{[m]}={\bf L}_{[m]}^{down}+{\bf L}_{[m]}^{up},\\
\eea
where ${\bf L}_{[m]}^{down}$ and  ${\bf L}_{[m]}^{up}$ are the down-Laplacian and the up-Laplacian of order $m$ and are defined as
\bea
{\bf L}_{[m]}^{down}&=&{\bf B}^{\top}_{[m]}{\bf B}_{[m]},\nonumber \\
{\bf L}_{[m]}^{up}&=&{\bf B}_{[m+1]}{\bf B}^{\top}_{[m+1]}.
\eea
The down-Laplacian ${\bf L}_{[m]}^{down}$ of order $m$, describes diffusion process taking place among $m$ simplices across $(m-1)$ shared simplices. For instance, the down-Laplacian of order $1$ describe diffusion from link to link across shared nodes.
The up-Laplacian ${\bf L}_{[m]}^{up}$ of order $m$ describes diffusion processes taking place among $m$ simplices across shared $(m+1)$ simplices. The up-Laplacian of order $1$ for example, describes the diffusion from link to link across shared triangles.

Interestingly the  spectral properties of the higher-order Laplacians can be proven to be independent on the orientation of the simplices as long as the orientation is induced by a labelling of the nodes.

One of the main results of Hodge theory  \cite{thesis,Barbarossa,High_Laplacian} is   that the degeneracy of the zero eigenvalues of the $m$- Laplacian ${\bf L}_{[m]}$ is equal to the Betti number $\beta_m$. The corresponding eigenvectors localize around the corresponding $m$-dimensional cavity of the simplicial complex.
It follows that if the simplicial complex has trivial topology, i.e. it is formed by a single connected component, $\beta_0=1$ and the simplicial complex has no higher-order cavities, (i.e.  $\beta_m=0$ for all $m>0$) then the $0$-Laplacian ${\bf L}_{[0]}$ has a zero eigenvalue that is not degenerate while all the higher-order Laplacians ${\bf L}_{[m]}$ with $m>0$ do not admit any zero eigenvalue. 

Let us observe here that Eq. (\ref{boundary2})  can be expressed in terms of the incidence matrices as 
\bea
{\bf B}_{[m-1]}{\bf B}_{[m]}&=&{\bf 0},\nonumber \\
{\bf B}^{\top}_{[m]}{\bf B}^{\top}_{[m-1]}&=&{\bf 0}.
\eea
From these relations it can be easily shown that the eigenvectors associated to   the non-null eigenvalues of ${\bf L}^{up}_{[m]}={\bf B}_{[m+1]}{\bf B}^{\top}_{[m+1]}$ are orthogonal to the eigenvectors associated with the non-null eigenvalues of 
${\bf L}^{down}_{[m]}={\bf B}^{\top}_{[m]}{\bf B}_{[m]}$.
Hodge theory therefore demonstrates (see for instance \cite{Barbarossa} for a gentle introduction) that the spectrum of the $m$-Laplacian includes all the non-null eigenvalues of the $m$-up-Laplacian and all the non-null eigenvalues of the $m$-down Laplacian. The   other eigenvalues of the $m$-Laplacian can only be zero and their degeneracy is given by the Betti number $\beta_{m}$.
Therefore the spectrum of the $m$-Laplacian is completely determined once the spectra of both the $m$-up-Laplacian and the $m$-down-Laplacian  are known.

Finally we observe that the up-Laplacians and the  down-Laplacians are related by  transposition
\bea
{\bf L}_{[m]}^{up}=[{\bf L}_{[m+1]}^{down}]^{\top}.
\eea
Therefore the spectrum of the $m$-up Laplacian is equal to the spectrum of the  $(m+1)$-down Laplacian.

Taking all these consideration together it follows that in order to know the spectrum of all higher-order Laplacians of a simplicial complex it is sufficient to know the spectrum of all its higher-order up-Laplacians.

Therefore in this work, without loss of generality we will focus on the spectral properties of $m$-up-Laplacians of pure $d$-dimensional simplicial complexes with order $0\leq m<d-1$.

\subsection{Up-Laplacians and their spectral dimension}

For a simplicial complex of dimension $d>m$ it is possible to define both a normalized and an un-normalized higher-order up $m$-Laplacian.
The un-normalized higher order up-Laplacian ${\bf {L}}^{up}_{[m]}={\bf B}_{[m+1]}{\bf B}^{\top}_{[m+1]}$ has elements 
\bea
[{L_{[m]}^{up}}]_{rq}=k_r^{[m]}\delta_{r,q}-\left(a^{[m]}_{\updownarrows}\right)_{rq}+\left(a_{\upuparrows}^{[m]}\right)_{rq},
\label{up1}
\eea
where ${\delta}_{x,y}$ indicates the Kronecker delta. In Eq. (\ref{up1})  we have used the {\em oriented upper incidence matrices} ${\bf a}^{[m]}_{\updownarrows}$ and ${\bf a}^{[m]}_{\upuparrows}$ defined as follows:  $(a_{\upuparrows}^{[m]})_{rq}=1$ if the two $m$-dimensional faces $r$ and $q$ are upper adjacent (they  are both incident to a $(m+1)$-dimensional simplex) and have similar orientation, otherwise $(a_{\upuparrows}^{[m]})_{rq}=0$; similarly   $(a_{\updownarrows}^{[m]})_{rq}=1$, if the two $m$-dimensional faces $r$ and $q$ are upper adjacent (they  are both incident to a $(m+1)$-dimensional simplex) and have dissimilar orientation,  otherwise $(a_{\updownarrows}^{[m]})_{rq}=0$. Finally  $k_r$ indicates the number of $(m+1)$-dimensional simplices incident to the $m$-dimensional simplex $r$.

 {The up-Laplacian can be used to characterize diffusion occurring among higher-order simplicies. In particular the   spectral properties of up-Laplacians  can affect the relaxation time of the diffusion process  as discussed in  Ref. \cite{High_Laplacian} for the simplicial complex model called ``Network Geometry with Flavor".}

Let us define the $N^{[m]}\times N^{[m]}$ matrix ${\bf K}_{[m]}$ as the diagonal matrix with diagonal elements $[{\bf K}_{[m]}]_{rr}=[L^{up}_{[m]}]_{rr}$. 
The normalized $m$-up-Laplacian $\hat{\bf L}_{[m]}^{up}$ can be defined as  
\bea
\hat{\bf L}_{[m]}^{up}={\bf K}_{[m]}^{-1/2}{\bf L}^{up}_{[m]}{\bf K}_{[m]}^{-1/2},
\label{norm}
\eea
where we note that in this expression we use the convention $0/0=0$.
The normalized $m$-up-Laplacian ${\hat{\bf L}}_{[m]}^{up}$ has  elements
\bea
[{\hat{L}_{[m]}}^{up}]_{rq}=\delta_{r,q}-\frac{1}{\sqrt{k_r^{[m]}k_q^{[m]}}}\left(a^{[m]}_{\updownarrows}\right)_{rq}+\frac{1}{\sqrt{k_r^{[m]}k_q^{[m]}}}\left(a_{\upuparrows}^{[m]}\right)_{rq}.
\eea
In this work we will focus on the spectral properties of the normalized up-Laplacians.
The spectrum  of the normalized and un-normalized $m$-up-Laplacians is in general distinct for simplicial complexes in which $k_r$ is dependent on $r$.
 However, we anticipate that when they both display a spectral dimension, their spectral dimension  is the same \cite{Burioni1}.

The density of eigenvalues $\bar{\rho}(\mu)$ of the normalized $m$-up-Laplacian  has a density of eigenvalues that includes a singular part formed by a delta function at $\mu=0$ and a regular part ${\rho}(\mu)$, i.e.
\bea
\bar{\rho}(\mu)=\bar{\rho}(0)\hat{\delta}(\mu)+{\rho}(\mu),
\eea
where we use $\hat{\delta}(x)$ to denote the delta function.
The emergence of the delta peak at $\mu=0$  can be easily explained.
First let us observe that Eq. (\ref{norm}) implies that the number of zero eigenvalues of the normalized and un-normalized $m$-up-Laplacians is the same.
Secondly let us note that the  spectrum of the  $m$-up-Laplacian  ${\bf L}^{up}_{[m]}$  can contain a highly degenerate zero eigenvalue. In fact, given the definition of the $m$-up-Laplacian   ${\bf L}^{up}_{[m]}={\bf B}_{[m+1]}{\bf B}^{\top}_{[m+1]}$ it follows that the eigenvalues  of the  $m$-up-Laplacian are the square of the singular values of the incidence matrix ${\bf B}_{[m+1]}$. Since the  incidence matrix ${\bf B}_{[m+1]}$ is a rectangular $N^{[m]}\times N^{[m+1]}$ matrix,  the non-zero singular values cannot be more than $\min(N^{[m]},N^{[m+1]})$. In particular for simplicial complexes with trivial topology,  the Hodge decomposition \cite{Barbarossa} implies  that the number  of non-zero eigenvalues of the $m$-up-Laplacian with $m>0$ are given by $\hat{N}^{[m]}=\min(N^{[m]},N^{[m+1]})$.  It follows that all the other eigenvalues are zero. Therefore for $m>0$ the degeneracy of the zero eigenvalue can be extensive, while for $m=0$ the degeneracy of the zero eigenvalue is given by the Betti number $\beta_0$, where $\beta_0=1$ for a trivial topology.
For a trivial topology the density of eigenvalues at $\mu=0$ of the graph Laplacian ($m$-up-Laplacian with $m=0$) is zero in the large network limit, while it can be greater than zero for $m>0$.

The normalized $m$-up Laplacian  displays a finite  {\it spectral dimension} $d_S$ when the regular part of its density of eigenvalues $\bar{\rho}(\mu)$ obeys the asymptotic behaviour
\bea
\rho(\mu)\cong C \mu^{d_S/2-1},
\label{scaling}
\eea
where $\mu\ll 1$ and $C$ is independent of $\mu$.

From this scaling  it directly follows that  the cumulative distribution ${\rho}_c(\mu)$ of the regular part of the density of eigenvalues ${\rho}(\mu)$, which is the integral of the  density of eigenvalues $0<\mu'\leq \mu$, follows the scaling 
\bea
{\rho}_c(\mu)\cong \tilde{C} \mu^{d_S/2}, 
\label{eq:rho_c}
\eea
for $\mu\ll 1$.  
This relation will prove  useful in the following, when we will numerically  compare the predicted spectral dimension with the numerical results.

\section{Simplicial complexes under consideration}
\subsection{Apollonian simplicial complexes of any dimension}

\begin{figure}[h!]
	\begin{center}
 \includegraphics[width=\columnwidth]{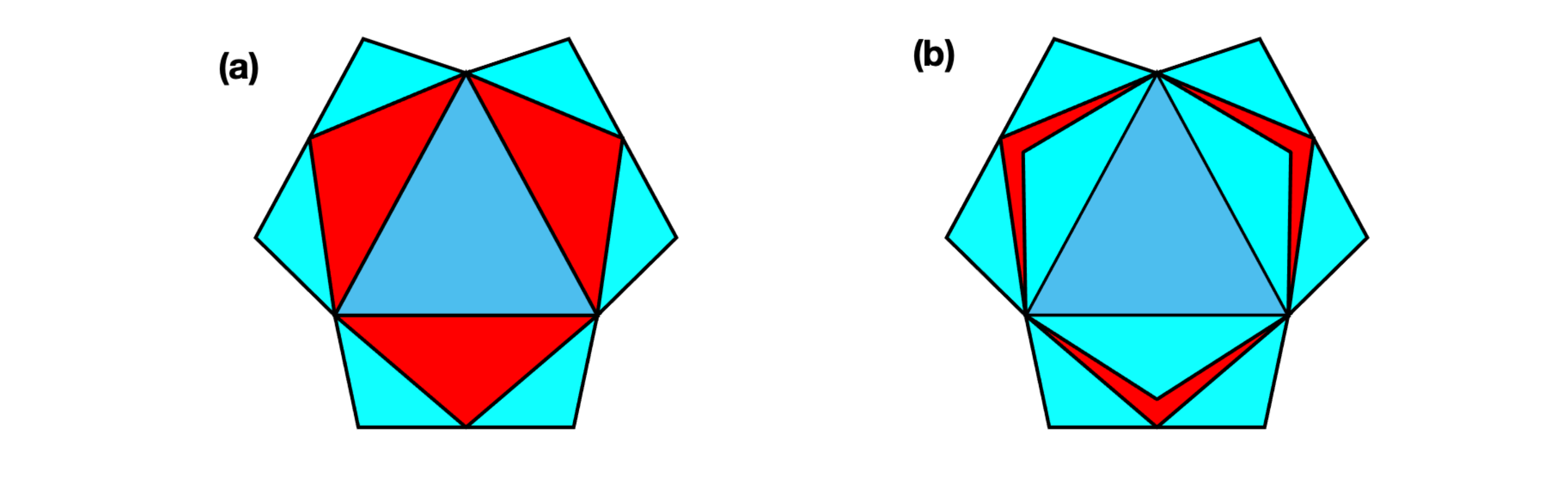}	
  \end{center}\caption{ {The $d=2$ dimensional Apollonian (panel (a)) and pseudo-fractal (panel (b)) simplicial complexes are shown at iteration $n=2$. The triangles added at iteration $n=0,1,2$ are shown in color blue, red and cyan respectively.}}
  	\label{fig:AppComplex}
	\end{figure}

A $d$-dimensional Apollonian simplicial complex \cite{apollonian,apollonian2} (with $d\geq 2$)  is generated iteratively by starting  from a single $d$-simplex at generation $n=0$ and adding a $d$-simplex at each generation $n>0$ to every $(d-1)$-dimensional face  introduced at the previous generation.  {In  Figure \ref{fig:AppComplex}a we show a $d=2$  dimensional Apollonian simplicial complex at iteration $n=2$.}

\subsection{Higher order Laplacian matrices of simplicial complexes}

At generation $n=0$ there are $N_0^{[m]}$ $m$-dimensional simplices in the simplicial complex with 
\bea
N_0^{[m]}={{d+1}\choose{m+1}}.
\eea
The number ${\mathcal N}_n^{[m]}$ of $m$-dimensional faces at generation $n$ is given by 
\bea
{\mathcal N}_n^{[m]} =(d+1)d^{n-1}  {d \choose m }
\label{NA}
\eea
In these Apollonian simplicial complexes, there are $N_n^{[m]}$ $m$-dimensional simplicial complexes at generation $n$ with 
\bea
N_n^{[m]}&=&N_0^{[m]}+\sum_{n'=1}^n{\mathcal N}_{n'}^{[m]}=(d+1)\left[\frac{d^n-1}{d-1}+\frac{1}{m-1}\right]{{d}\choose m},
\label{NA0}
\eea
Finally we note here that in the following we will used the notation $Q^{[m]}$ to indicate  the set of $m$-simplices of the Apollonian simplicial complex.

The Apollonian simplicial complex are small-world, i.e. their skeleton has an infinite Hausdorff dimension,
\bea
d_H=\infty,
\eea
therefore at each generation their diameter  grows logarithmically with the total number of nodes of the network.
Moreover, the Apollonian simplicial complex of dimension $d$ are manifolds that define discrete hyperbolic lattices including for $d=2$ the Farey graph.

Let us add here a pair of additional combinatorial properties of Apollonian simplicial complex that will be useful later.
At each generation $n$ we call {\em simplices  of type} $\ell$ the simplices added at generation $n^{\prime}=n-\ell$.
At generation $n$, the number of $d$-simplices of  generation $n$ attached to simplices of dimension $m$   (with $m<d$)  of type $\ell>0$ is given by
\bea
w_{\ell}^{[m]}=(d-m)(d-m-1)^{\ell-1}.
\label{wl}
\eea
Moreover, we observe that the  number of $(m+1)$-dimensional simplices of generation $n$ incident to  $m$-simplices  added  at generation $n^{\prime}=n-\ell$ is given by 
$w_{\ell}^{[m]}$ for $\ell>0$ and  $w_{\ell}^{[m]}=d-m$ for $\ell=0$.

\subsection{Pseudo-fractal simplicial complexes of any dimension}

A pseudo-fractal simplicial complex \cite{pseudo_Doro} of dimension $d$ with $d\geq 2$ is constructed iteratively. At generation $n=0$ the simplicial complex is formed by a  single $d$-simplex (with $d\geq 2$).
At each generation $n>0$ we glue a $d$-simplex to every $(d-1)$-dimensional face introduced at generation $n\geq 0$.  {In Figure $\ref{fig:AppComplex}$b we show a  of $d=2$ dimensional pseudo-fractal simplicial complex at iteration $n=2$. }
At generation $n=0$ the  number of $m$-dimensional simplices  ${N}_0^{[m]}$ is given by 
\bea
N_0^{[m]}={{d+1}\choose{m+1}}.
\eea 
The number ${\mathcal N}_n^{[m]}$ of $m$-dimensional faces added at generation $n>0$ is given by
\bea
{\mathcal N}_n^{[m]}=(d+1)^n{d\choose{m}}.
\label{NP}
\eea
The number $m$-dimensional faces  $N_n^{[m]}$  at generation $n$ is 
\bea
N^{[m]}_n&=&N_0^{[m]}+\sum_{n'=1}^n{\mathcal N}_{n'}^{[m]}\nonumber \\
&=&(d+1)\left[\frac{(d+1)^n-1}{d}+\frac{1}{m+1}\right]{d\choose{m}}.
\label{NP0}
\eea
The pseudo-fractal simplicial complexes differs from Apollonian simplicial complexes significantly as they are not discrete manifolds.
However both simplicial complexes have an underlying non-amenable network structure and are characterized by having a  small Cheeger constant.

Moreover the  pseudo-fractal simplicial complexes, as the Apollonian simplicial complexes,  have a   small-world skeleton, i.e. their underlying networks have  an infinite Hausdorff dimension  
\bea
d_H=\infty.
\eea

For pseudo-fractal simplicial complexes we use the same notation as for Apollonian simplicial complex and we indicate $Q^{[m]}$  the set of $m$-simplices of the pseudo-fractal simplicial complex.
Additionally  we indicate  as  {\em simplices of type} $\ell$ the simplices added at generation $n^{\prime}=n-\ell$, in the pseudo-fractal simplicial complex evolved up to   generation $n$.
We make the following useful remark: at generation  $n$ the number of $d$-simplices of   generation $n$ attached to $m$-simplices (with $m<d$) of type $\ell>0$ is given by
\bea
\hat{w}_{\ell}^{[m]}=(d-m)\sum_{\ell^{\prime}=0}^{\ell}(d-m-1)^{\ell^{\prime}-1}.
\eea
Finally the number of $(m+1)$-simplices of generation $n$ added to $m$-simplices of generation $n^{\prime}=n-\ell$ is given by $\hat{w}_{\ell}^{[m]}$ for $\ell>0$ and $\hat{w}_0^{[m]}=(d-m)$ for $\ell=0$.

\section{Gaussian model and the RG approach}
\subsection{The ensemble of weighted normalized Laplacians}
In this section our goal is to define the theoretical framework of a real space RG approach to calculate the spectrum of the  normalized  $m$-dimensional up-Laplacian  of the Apollonian  and the pseudo-fractal simplicial complexes.
The renormalization group  acts on a weighted simplicial complex in which we attribute a weight $p_{\tau}$ to each $(m+1)$-dimensional simplex $\tau$ while the topology of the simplicial complex remains fixed.
Therefore in the RG approach we investigate the  RG flow defined over the ensemble of   weighted normalized up-Laplacian matrices ${\bf \hat{L}}_{[m]}^{up}$ of elements
\bea
[\hat{L}_{[m]}^{up}]_{rq}=\delta_{r,q}-\frac{1}{\sqrt{s_r^{[m]}s_q^{[m]}}}p_{\tau(r,q)}\left(a^{[m]}_{\updownarrows}\right)_{rq}+\frac{1}{\sqrt{s_r^{[m]}s_q^{[m]}}}p_{\tau(r,q)}\left(a_{\upuparrows}^{[m]}\right)_{rq},
\eea
where $p_{\tau(qr)}$ indicates the weight of the $(m+1)$-dimensional simplex $\tau$ incident to both $r$ and $q$   and $s_r$ indicates the strength of the simplex  $r$, i.e. $s_r=\sum_{\tau\supset r}^{N_{n}^{[m+1]}} p_{\tau}$.
From here on, we will focus on finding the density of eigenvalues of the up-Laplacian of order $m$. In the following sections we will therefore adopt a simplified notation, dropping the indication "up" and the index $[m]$ in most of the relevant mathematical quantities. We will therefore indicate ${\bf \hat{L}}_{[m]}^{up}$ simply as ${\bf \hat{L}}$, $N_n^{[m]}$ as $N_n$, $a^{[m]}_{\updownarrows}$ as $a_{\updownarrows}$, $a_{\upuparrows}^{[m]}$ as $a_{\upuparrows}$ and so on.

\subsection{Gaussian models and Laplacian spectrum}

The density of eigenvalues of a symmetric matrix can be derived analytically using the properties of the Gaussian model following a standard procedure of statistical mechanics  \cite{Kahng} quite common in Random Matrix Theory \cite{Mehta,RandomMatrix}.
Therefore if we want to derive the density of eigenvalues of the   $m$-dimensional up-Laplacian $\hat{{\bf L}}$ which for generation $n$ will be a $N_n\times N_n$ symmetric matrix we should consider  the Gaussian model whose  partition function  reads
\bea
Z(\mu)=\int {\mathcal D}\psi \exp\left[i\mu \sum_{r}\psi_r^2-i\sum_{rq}\hat{L}_{rq}\psi_r\psi_q\right]=\frac{(i\pi)^{N_n/2}}{\sqrt{\prod_{r}(\mu-\mu_r)}}
,
\label{z}
\eea
where $\mu_r$ are the eigenvalues of the normalized up-Laplacian matrix $\hat{\bf L}$ and the differential ${\mathcal D}\psi$ stands for 
\bea
{\mathcal D}\psi=\prod_{r=1}^{N_n}\left(\frac{d\psi_r}{\sqrt{2\pi}}\right).
\eea  

By changing variables and putting $\phi=\psi/\sqrt{s_r}$ the  partition function can  be  rewritten as 
\bea
Z(\mu)=\prod_{r=1}^{N_n}\sqrt{s_r}\int {\mathcal D}\phi e^{iH(\{\phi\})},
\eea
with 
\bea
H(\{\phi\})=\sum_{\tau\in Q^{[m+1]}}p_\tau\left[-(1-\mu)\sum_{r\subset \tau} \phi_{r}^2+2\sum_{r<q|r,q\subset t}(a^{\updownarrows}_{rq}-a^{\upuparrows}_{rq})\phi_r\phi_q\right],
\eea
where  $r,q$ are both $m$-simplices, i.e. $r,q\in Q^{[m]}$.
The spectral density $\bar{\rho}(\mu)$ of the normalized Laplacian matrix can be found using the relation 
\bea
\bar{\rho}(\mu)=-\frac{2}{\pi}\mbox{Im} \frac{\partial f}{\partial \mu},
\label{rm0}
\eea 
where $f$ is the free-energy density  defined as 
\bea
f=-\lim_{n\to\infty}\frac{1}{N_n}\ln Z(\mu).
\label{rm}
\eea
In fact, inserting   Eq. (\ref{z}) in the Eq. (\ref{rm}) we obtain
\bea
f=-\lim_{n\to \infty}\left[\frac{1}{N_n}\sum_{r=1}^N\frac{1}{2}\ln(\mu-\mu_r)\right]-\frac{1}{2}\ln (i\pi).
\eea
Therefore we can show that Eq. (\ref{rm0}) is correct by plugging the final expression for $f$ in Eq.(\ref{rm0}),
\bea
\hspace{-5mm}\bar{\rho}(\mu)=-\frac{2}{\pi}\mbox{Im} \frac{\partial f}{\partial \mu}=\frac{1}{\pi}\lim_{n\to \infty}\frac{1}{N_n}\mbox{Im} \sum_{r=1}^{N_n}\frac{1}{\mu-\mu_r}=\lim_{n\to \infty}\frac{1}{N_n} \sum_{r=1}^{N_n}\hat{\delta}({\mu-\mu_r}).
\eea

\subsection{The general RG approach}

As was the case in Ref.\cite{RG1}, where the spectrum of the $0$-Laplacian was derived using the RG flow, the   parameters $p$ and $\mu$ are renormalized differently for faces  of different type $\ell$ when we study the spectrum of the $m$-dimensional up-Laplacian. The partition function $Z_n(\bm{\omega})$ corresponding to the Gaussian model of the simplicial complex evolved up to generation $n$ is a function of the parameters $\bm{\omega}=(\{\mu_{\ell}\},\{p_{\ell}\})$, and can be expressed as 
\bea
 Z_n(\bm{\omega})=\int \mathcal{D}\phi e^{iH(\{\phi\})},
 \eea
where 
\bea
\hspace{-20mm}H(\{\phi\})&=&\sum_{\ell=0}^{n}\sum_{\tau\in Q_n^{[m+1]}(\ell)}\left[-i(1-\mu_{\ell})p_{\ell}\sum_{r\subset \tau}\phi_r^2+2ip_{\ell}\sum_{r<q|r\subset \tau,q\subset \tau}(a_{rq}^{\updownarrows}-a_{rq}^{\upuparrows})\phi_r\phi_q\right],\label{H0}
\eea
with $Q_n^{[m+1]}{(\ell)}$ indicating the set of $(m+1)$-dimensional simplices of type $\ell$ in a simplicial complex evolved up to generation $n$ and with $r,q\in Q^{[m]}$.
The Gibbs measure of this Gaussian model is given in terms of the Hamiltonian $H(\{\phi\})$ defined in Eq. (\ref{H0}) as
\bea
P_n(\{\bm{\phi}\})=\frac{1}{Z(\bm{\omega})}e^{iH(\{\phi\})}.
\label{Gibbs}
\eea
In order to calculate the partition function $ Z_n(\bm{\omega})$ we adopt a real space renormalization group approach. We will first integrate the Gaussian fields corresponding to the ${\mathcal N_n}$ $m$-dimensional simplices added to the simplicial complex at generation $n$ and then iteratively integrate over the simplices added at generation $n-1$ and so forth, until all the integrals in the definition of the partition function $Z_n(\bm{\omega})$ are performed.
More specifically we  consider  the following real space renormalization group procedure. We start with initial conditions $\mu_\ell=\mu$ and $p_{\ell}=1$ for all values of $\ell>0$. At each RG iteration, we integrate over the Gaussian variables $\phi_{\bar{r}}$ associated to simplices $\bar{r}\in {\mathcal N_n}$ and we rescale the remaining Gaussian variables in order to obtain the  renormalized Gibbs measure $P(\{\bm{\phi'}\})$ of the same type as Eq.~(\ref{Gibbs}) but with rescaled parameters $(\{\mu_{\ell}^{\prime}\},\{p_{\ell}^{\prime}\})$, i.e. 
\bea
P_{n-1}(\{\bm{\phi'}\})=\left.\int {\mathcal D}\phi^{(n)} P_n(\{\bm{\phi}\})\right|_{\bm{\phi'}={\bf F}(\{\bm{\phi}\})},
\eea
where
\bea
{\mathcal D}\phi^{(n)}=\prod_{r\in {\mathcal N_n}}\left(\frac{d\phi_r}{\sqrt{2\pi}}\right).
\eea 
The  fields are rescaled in a way that keeps $p_{1}=1$ at each iteration of the RG flow, i.e. the weight of the $(m+1)$-dimensional faces of type $\ell=1$ is always fixed to one.
It follows that at each step of the RG transformation we have 
\bea
H(\{\phi\})\to H^{\prime}(\{\phi^{\prime}\}),
\eea
where, 
\bea
\hspace{-25mm}H^{\prime}(\{\phi\})=\sum_{\ell=1}^{n-1}\sum_{\tau\in Q_{n-1}^{[m+1]}{(\ell)}}\left\{-(1-\mu_{\ell}^{\prime})p_{\ell}^{\prime}\sum_{r\subset \tau}(\phi_r^{\prime})^2+2p_{\ell}^{\prime}\sum_{r<q|r,q\subset \tau}[a_{rq}^{\updownarrows}-a_{rq}^{\upuparrows}]\phi_r^{\prime}\phi_q\right\}.
\eea
This procedure allows us to determine  the  renormalization group transformation $R$ acting on the model parameters ${\bm{\omega}}=(\{\mu_{\ell}\},\{p_{\ell}\})$,
\bea
\bm{\omega}'=R \bm{\omega}.
\eea
Under the renormalization group flow, the partition function  transforms  according to   
\bea
Z_{n}(\bm{\omega})=e^{-{N}_n g(\bm{\omega})}Z_{n-1}(\bm{\omega}').
\label{ZRG}
\eea

By using Eq. (\ref{NA}) and Eq. (\ref{NP}), the free energy density at generation $n$
\bea
f&=&-\lim_{n\to \infty}\frac{1}{N_n}\ln Z_n(\bm{\omega})
\label{fg}
\eea
can be approximated as
\bea
f&\simeq&\sum_{\tau=0}^{\infty}\frac{g(R^{(\tau)}\bm{\omega})}{d^{\tau}},
\label{fg1}
\eea
for the Apollonian simplicial complexes, and  as 
\bea
f&\simeq&\sum_{\tau=0}^{\infty}\frac{g(R^{(\tau)}\bm{\omega})}{(d+1)^{\tau}},
\label{fg2}
\eea
for the pseudo-fractal simplicial complexes.

We will show in the next section that the RG flow for this Gaussian model is determined by the fixed point at $\mu^{\star}=0$. This implies that the spectral dimension of higher-order up-Laplacians is universal  \cite{Burioni1}, i.e. it is the same for normalized and un-normalized up-Laplacians.

\section{General RG equations for the Apollonian simplicial complex}

\subsection{The integral}
To derive the renormalization group equations for the Apollonian simplicial complex we need to  perform the integration  over the Gaussian fields  associated to the $m$-simplices  added at generation $n$.  In the Apollonian simplicial complex, any $d$-simplex of generation $n$ is only incident to $d$-simplices added at previous generations. Specifically, every new $d$-simplex contains a single new node and shares exactly one of its $(d-1)$-faces with the Apollonian simplicial complex at the previous iteration. Therefore the integrations over all $m$-simplices  added at iteration $n$ can be performed independently by separately considering the Gaussian fields corresponding to $m$-simplices belonging to different $d$-simplices added at iteration $n$. Consequently in this paragraph we only focus on the integration over the Gaussian fields associated to $m$-simplices belonging to a single $d$-simplex of generation $n$. \\

 In order to perform this integral let us define some notation.
 Given the  generic $d$-simplex $\bar{r}$ added at iteration $n$, i.e. $\bar{r}\in {\mathcal N}_n$, we indicate with $j$ its most recent node, i.e the single  node $j\subset \bar{r}$ of type  $\ell=0$.  Each $d$-simplex $\bar{r}$ added at generation $n$ contains ${{d}\choose{m}}$ new $m$-simplices  added at generation $n$. All these simplices   include the node $j$ and $m$ other nodes out of the $d$ nodes  of type $\ell>0$ belonging to $\bar{r}$.  We will denote the set of these $m$-simplices by ${\mathcal M}_n$ and the Gaussian fields associated to the $m$-simplices $q\in {\mathcal M}_n$ by $\bar{\psi}_q$.
Additionally, the simplex $\bar{r}$ contains ${d}\choose{ m+1}$ $m$-faces formed exclusively by nodes of type $\ell>0$. 
We will denote the set of these simplices by ${\mathcal R}_n$ and the Gaussian fields associated to the $m$-simplices $q\in {\mathcal R}_n$ by ${\phi}_q$.
Finally, let us define ${\mathcal Q}^{[m+1]}$ to be the set of all $(m+1)$-dimensional faces of the simplex $\bar{r}$ added at iteration $n$.
With this notation, the integral over the fields $\{\bar{\psi}_{\bar{r}}\}$ reads,
\bea
I_{m}=\int\mathcal{D}\bar{\psi} \ \exp\left\{i\left[H_0(\{\bar{\psi}\},\{\phi\})+H_1(\{\bar{\psi}\},\{\phi\})\right]\right\},
\label{dIntegral}
\eea
where  $H_0(\{\bar{\psi}\},\{\phi\})$ is given by
\bea
H_0(\{\bar{\psi}\},\{\phi\})&=&-(1-\mu_1)\left[(d-m) \sum_{q\in {\mathcal M}_n} \bar{\psi}_q^2+\sum_{q\in {\mathcal R}_n}\phi_q^2\right], 
\label{dHamiltonian1}
\eea

and $H_1(\{\bar{\psi}\},\{\phi\})$ is given by
\bea
\hspace{-10mm}H_1(\{\bar{\psi}\},(\{\phi\})=2\sum_{\tau\in {\mathcal Q}^{[m+1]}}\left[ \sum_{r<q|r \subset \tau, q\subset \tau}A_{rq}\bar{\psi}_{r} \bar{\psi}_{q}+\sum_{r,q|r\subset\tau, q\subset \tau}A_{rq}\bar{\psi}_{r} \phi_{q}\right].
\label{dHamiltonian2}
\eea

Here $A_{rq}$ is given by
\bea
A_{rq}=a_{rq}^{\updownarrows}-a_{rq}^{\upuparrows},
\eea
and ${\mathcal D}\bar{\psi}$ is defined by
\bea
{\mathcal D}\bar{\psi}=\prod_{q\in { \mathcal M}_n}\left(\frac{d\bar{\psi}_q}{\sqrt{2\pi}}\right).
\eea
The integral $I_{m}$ is given by
\bea
&&\hspace{-25mm}I_{m}=\exp\left\{-{i}(1-\mu_1)\sum_{r\in {\mathcal M}_{n}}\phi_{r}^2+\frac{i}{d-(d-m)\mu_1}\left[(m+1)\sum_{r\in {\mathcal M}_n}\phi_{r}^2+2\sum_{r<q}A_{rq}\phi_r\phi_q)\right]\right\}\nonumber \\
&&\times (-i)^{{d\choose m}/2}(-1)^{-{{d-1}\choose {m-1}}/2}\pi^{{d\choose m}/2}(d-m)^{{d-1}\choose{m-1}}G(\mu_1)^{-1/2},
\label{I}
\eea
where
\bea
G(\mu_1)=[d-(d-m)\mu_1]^{{d-1}\choose {m}}\mu_1^{{d-1}\choose {m-1}}.
\label{G}
\eea
We note that for $m=d-1$, the cardinality of the set ${\mathcal M}_n$ equals one. Therefore  the integral $I_{d-1}$ simplifies to
\bea
&&\hspace{-15mm}I_{d-1}=\exp\left\{{i}\left[-(1-\mu_1)+\frac{d}{d-\mu_1}\right]\sum_{r\in {\mathcal M}_{n}}\phi_{r}^2\right\} (-i\pi)^{d/2}(i)^{-{({d-1})}}G(\mu_1)^{-1/2},
\label{Idm1}
\eea
and $G(\mu_1)$ given by Eq.(\ref{G}) simplifies to
\bea
G(\mu_1)=(d-\mu_1)\mu_1^{{d-1}}.
\label{G_d_1}
\eea
Given the different structure of the integral $I_{m}$ for $m\leq d-2$ and for $m=d-1$, we will treat the case $m\leq d-2$ and the case $m=d-1$ separately in the subsequent paragraphs.
\subsection{The RG equations for $m\leq d-2$}
In this section we will show that the  RG equations 
\bea
\bm{\omega}'=R \bm{\omega}.
\eea
for the Apollonian simplicial complex for $m\leq d-2$ have the explicit expression,
\bea
&&\hspace{-20mm}(1-\mu_{\ell}')p_{\ell}^{\prime}=\left((1-\mu_1)(d-m-2)^{\ell-1}+(1-\mu_{\ell+1})p_{\ell+1}-\frac{(m+1)(d-m-2)^{\ell-1}}{d-(d-m)\mu_1}\right)\nonumber \\
&&\times \left[p_2+\frac{(d-m-1)}{d-(d-m)\mu_1)}\right]^{-1},\nonumber \\
&&\hspace{-20mm}p'_{\ell}=\left[p_{\ell+1}+\frac{(d-m-1)}{d-(d-m)\mu_1}(d-m-2)^{\ell-1}\right]\left[p_2+\frac{(d-m-1)}{d-(d-m)\mu_1}\right]^{-1},
\label{RGhdA}
\eea
for all $\ell\geq 1$. The initial conditions for all $\ell\geq 1$ are $(\mu_{\ell},p_{\ell})=(\mu,1)$ with $\mu\ll 1$.
This result generalizes the RG equations that were found in Ref.\cite{RG1} and can be derived using a similar procedure. The results derived in Ref.\cite{RG1} correspond to the case of $m=0$ in Eqs. (\ref{RGhdA}).

According to the renormalization group procedure explained in the previous section, we have to integrate over each $m$ simplex  $\bar{r}\in {\mathcal N}_n$ at each iteration of the RG procedure.
Each integration over the generic simplex $\bar{r}$ performed in Eq.$(\ref{I})$ contributes to the Hamiltonian $H^{\prime}(\{\phi^{\prime}\})$ with a term
\bea
\hspace{-15mm}-(1-\mu_1)\sum_{q}\phi_{q}^2+\frac{1}{d-(d-m)\mu_1}\left[(m+1)\sum_{q}\phi_{q}^2+2\sum_{r<q}A_{rq}\phi_r \phi_q)\right].
\label{IH}
\eea
If we just focus on the term coupling different Gaussian fields for any ${(m+1)}$-dimensional simplex which include both $q$ and $r$ the contribution is,
\bea
\left[\left(2\frac{1}{d-(d-m)\mu_1}\right)A_{rq}\phi_r \phi_q\right].
\eea
In the Apollonian simplicial complex, there are  $w_{\ell}^{[m+1]}$ $d$-simplices of iteration $n$ incident to a $(m+1)$-simplex of type $\ell$, including both the $m$ simplex $q$ and simplex $r$. The overall contribution to the term  proportional to $\phi_r \phi_q$  in $H^{\prime}(\{\phi^{\prime}\})$ is 
\bea
\left[\left(2\frac{1}{d-(d-m)\mu_1}\right)w_{\ell}^{[m+1]}A_{rq}\phi_r \phi_q\right].
\eea
It follows that, before rescaling, the overall contribution of the integrals over $\bar{r}\in {\mathcal N}_n$  to the term of the  Hamiltonian $H^{\prime}(\{\phi^{\prime}\})$ proportional to $\phi_r \phi_q$ is given by
\bea
\left\{2\left[p_{\ell+1}+\left(\frac{1}{d-(d-m)\mu_1}\right)w_{\ell}^{[m+1]}\right]A_{rq}\phi_r \phi_q\right\}.
\eea
The real space RG procedure prescribes that after rescaling of the fields $\phi_q\to \phi^{\prime}_q$, we should have
\bea
\left\{2\left[p_{\ell+1}+\left(\frac{1}{d-(d-m)\mu_1}\right)w_{\ell}^{[m+1]}\right]A_{rq}\phi_r \phi_q\right\}=\left\{2p_{\ell}^{\prime}A_{rq}\phi^{\prime}_r \phi_q^{\prime}\right\}.
\eea
The correct rescaling of the fields that ensures $p_1^{\prime}=p_1=1$ is given by
\bea
\phi'&=&\phi\left[p_2+\frac{d-m-1}{d-(d-m)\mu_1}\right]^{1/2}.
\label{rescaling}
\eea
Here we have used $w_{1}^{[m+1]}=(d-m-1)$.
Finally, by using Eq. (\ref{wl}) for $w_{\ell}^{[m+1]}$, the RG equation for $p^{\prime}_{\ell}$ reads 
\bea
\hspace{-5mm}p'_{\ell}=\left[p_{\ell+1}+\frac{(d-m-1)(d-m-2)^{\ell-1}}{d-(d-m)\mu_1}\right]\left[p_2+\frac{d-m-1}{d-(d-m)\mu_1}\right]^{-1}.
\eea
In order to find the RG equations for  $\mu_{\ell}^{\prime}$, we need to consider the contribution  to the rescaled Hamiltonian coming from the integral $I_m$ in Eq. (\ref{IH}) that is  proportional to $\phi_q^2$. This contribution is,
\bea
\left[\left(-(1-\mu_1)+\frac{m+1}{d-(d-m)\mu_1}\right)\phi_q^2\right].
\eea
Since there are ${w}_{\ell}^{[m]}$ $d$-simplicies of generation $n$ incident to the $m$-simplex $q$ added at generation $n^{\prime}=n-\ell$, the integration over the Gaussian fields corresponding to the simplices added at generation $n$ contributes,
\bea
\left[\left(-(1-\mu_1)+\frac{m+1}{d-(d-m)\mu_1}\right){w}_{\ell}^{[m]}\phi_q^2\right].
\eea
to the Hamiltonian for each $m$-dimensional simplex $q$.
Let us now equate the term proportional to  $\phi_q^2$ in  the Hamiltonian before and after the rescaling of the fields, i.e. 
\bea
&&\hspace{-15mm}\left\{\left[-\sum_{\ell'=1}^{\ell}(1-\mu_{\ell^{\prime}+1})p_{\ell^{\prime}+1}w^{[m]}_{\ell-\ell^{\prime}}+\left(-(1-\mu_1)+\frac{m+1}{d-(d-m)\mu_1}\right){w}_{\ell}^{[m]}\right](\phi_q)^2\right\}\nonumber \\
&=&\left\{\left[-\sum_{\ell'=1}^{\ell}(1-\mu^{\prime}_{\ell^{\prime}})p_{\ell^{\prime}}^{\prime}w_{\ell-\ell^{\prime}}^{[m]}\right](\phi^{\prime}_q)^2\right\}.
\label{xprime}
\eea
We observe that the coefficients $w_{\ell}^{[m]}$ can be written as
\bea
w_{\ell}^{[m]}=\sum_{\ell^{\prime}=1}^{\ell}w^{[m]}_{\ell-\ell^{\prime}}c_{\ell^{\prime}},
\label{combinatorial}
\eea
where $c_{\ell^{\prime}}$ is given by 
\bea
c_{\ell}=(d-m-2)^{\ell-1}.
\eea

After rescaling the fields according to Eq.~(\ref{rescaling}), using Eq. (\ref{combinatorial}) and Eq. (\ref{xprime}) we get the RG equation for $\mu_{\ell}$,
\bea
\hspace{-25mm}(1-\mu_{\ell}')p_{\ell}^{\prime}&=&\left((1-\mu_1)(d-m-2)^{\ell-1}+(1-\mu_{\ell+1})p_{\ell+1}-\frac{(m+1)(d-m-2)^{\ell-1}}{d-(d-m)\mu_1}\right)\nonumber \\
&&\times \left[p_2+\frac{d-m-1}{d-(d-m)\mu_1}\right]^{-1}.
\eea

This completes our derivation of the RG equations Eq.(\ref{RGhdA}).

\subsection{The free-energy density and spectral dimension for $m\leq d-2$}

Using the renormalization group and in particular equation Eq. (\ref{ZRG}) for the partition function, we can calculate the function $g(\bm{\omega})$
\bea
g(\bm{\omega})=\frac{{\mathcal N}_n}{2N_n}\ln G(\mu_1)+\frac{N_{n-1}}{2N_{n}}\ln \left[p_2+\frac{(d-m-1)}{ {d}-(d-m)\mu_1}\right]+c,
\label{gw}
\eea
where $c$ indicates a constant. The first term on the right hand side of this equation comes from the result of the integral $I_m$ in Eq. (\ref{I}). The second term is the contribution due to the rescaling of the fields given by Eq. (\ref{rescaling}). 
Given this expression for $g(\bm{\omega})$ the free energy density $f$ can be obtained from Eq. (\ref{fg1}),
\bea
f&\simeq&\sum_{\tau=0}^{\infty}\frac{g(R^{(\tau)}\bm{\omega})}{d^{\tau}}\nonumber \\ &&\hspace{-6mm}\simeq \sum_{\tau=0}^{\infty}\frac{1}{d^{\tau}}\left\{
\frac{(d-1)}{2d}\ln G\left(\mu_1^{(\tau)}\right)+\frac{1}{2d}\ln \left[p_2^{(\tau)}+\frac{d-m-1}{d-(d-m)\mu_1^{(\tau)}}\right]\right\}.
\eea
Anticipating that the relevant fixed point at $(\mu^{\star},p_2^{\star})=(0,p^{\star})$ is repulsive, we  assume that close to this fixed point the RG flow can be described by the equations 
\bea
\mu_1^{(\tau)}\simeq\mu\lambda^{\tau}\nonumber \\
p_2^{(\tau)}\simeq p^{\star}+\lambda^{\tau}(1-p^{\star})
\label{gflow1}
\eea
where $\mu_1^{(\tau)}$ and $p_2^{(\tau)}$ indicate  the value of $\mu_1$ and $p_2$ at the iteration $\tau$ of the RG transformation, and where  $\lambda>1$ is the largest eigenvalue of the RG equations linearlised close to the relevant fixed point.
Therefore using Eq. (\ref{rm0}) the spectral density $\bar{\rho}(\mu)$ can be found by,
\bea
&&\hspace{-20mm}\bar{\rho}(\mu)\simeq \frac{2}{\pi}\mbox{Im}\sum_{\tau=0}^{\infty}\frac{1}{d^{\tau}}\frac{\partial g(\mu_1^{(\tau)},p_2^{\tau})}{\partial \mu}\nonumber \\&&\hspace{-20mm}\simeq \frac{2}{\pi}\mbox{Im}\sum_{\tau=0}^{\infty}\frac{\lambda^{\tau}}{d^{\tau}}\frac{(d-1)}{2d}\left[{{d-1}\choose{m}}\frac{1}{d-(d-m)\mu_1^{(\tau)}}+{{d-1}\choose{m-1}}\frac{1}{\mu_1^{(\tau)}}\right]\nonumber \\
&&\hspace{-20mm}+ \frac{2}{\pi}\mbox{Im}\sum_{\tau=0}^{\infty}\frac{\lambda^{\tau}}{d^{\tau}}\frac{d-m}{2d}y\left[\left(p_2^{(\tau)}\left[d-(d-m)\mu_1^{(\tau)}\right]+y\right)\left(d-(d-m)\mu_1^{(\tau)}\right)\right]^{-1},\nonumber 
\eea
where $y=d-m-1$.
We notice that for $m>0$ the spectrum acquires a delta peak at $\mu=0$, corresponding to the finite density of zero eigenvalues of the up-Laplacian, i.e.
\bea
\bar{\rho}(\mu)=\bar{\rho}(0)\hat{\delta}(\mu)+{\rho}(\mu).
\label{rho_delta}
\eea
In fact by using the relation 
\bea
\frac{1}{\pi}\mbox{Im}\frac{1}{\mu}=\hat{\delta}(\mu),
\eea
and the RG flow given by Eq. (\ref{gflow1}), we have 
\bea
\frac{2}{\pi}\mbox{Im}\sum_{\tau=0}^{\infty}\frac{\lambda^{\tau}}{d^{\tau}}\frac{d-1}{2d}{{d-1}\choose{m-1}}\frac{1}{\mu_1^{(\tau)}}=\bar{\rho}(0)\hat{\delta}(\mu),
\eea
where 
\bea
\bar{\rho}(0)=\frac{d-1}{d}{{d-1}\choose{m-1}}\frac{1}{1-1/d}.
\eea
The regular part of the density of eigenvalues ${\rho}(\mu)$  is given by 
\bea
&&\hspace{-26mm}{\rho}(\mu)\simeq \frac{2}{\pi}\mbox{Im}\sum_{\tau=0}^{\infty}\frac{\lambda^{\tau}}{d^{\tau}}\frac{(d-1)}{2d}\left[{{d-1}\choose{m}}\frac{1}{d-(d-m)\mu_1^{(\tau)}}\right]\nonumber \\
&&\hspace{-26mm}+ \frac{2}{\pi}\mbox{Im}\sum_{\tau=0}^{\infty}\frac{\lambda^{\tau}}{d^{\tau}}\frac{d-m}{2d}y\left[\left(p^{(\tau)}\left[d-(d-m)\mu_1^{(\tau)}\right]+y\right)\left(d-(d-m)\mu_1^{(\tau)}\right)\right]^{-1}.
\label{bar_rhoe}
\eea
This expression can be approximated by substituting the sum over $\tau$ with an integral. Upon changing the variable of this integral to   $z=\lambda^{\tau}$ we can  use the theorem of residues at $\mu_1^{(\tau)}=z\mu=d/(d-m)$ to solve the integral, obtaining  the asymptotic scaling 
\bea
{\rho}(\mu)&\simeq& C \mu^{d_S/2-1},
\eea
where the spectral dimension $d_S$ is given by,
\bea
d_S=2\frac{\ln d}{\ln \lambda}.
\label{ds0}
\eea
Note however, that Eq. (\ref{ds0}) holds only if the RG flow can be approximated by Eq.(\ref{gflow1}) for $\mu_1^{(\tau)}\simeq d/(d-m)$.

\subsection{RG equations  for $m=d-1$}
In this paragraph we will show that for $m=d-1$, the RG equations read,
 \bea
&&p_{\ell}=p_1=1\nonumber \\
&&(1-\mu_{\ell}^{\prime})=(1-\mu_{\ell+1})+(-1)^{\ell}\left[(1-\mu_1)-\frac{d}{d-\mu_1}\right],
\label{RG2A}
\eea
for all $\ell\geq 1$ with initial conditions  $(\mu_{\ell},p_{\ell})=(\mu,1)$ with $\mu\ll 1$.

First we observe that for  $m=d-1$ the contribution of the integral $I_{d-1}$ to the Hamiltonian $H^{\prime}(\phi^{\prime})$ is given by \bea
\left\{\left[-(1-\mu_1)+\frac{d}{d-\mu_1}\right]\sum_{r\in {\mathcal M}_{n}}\phi_{r}^2\right\}.
\eea This contribution does not contain  any term proportional to $\phi_r \phi_q$. This observation  automatically indicates that $p_{\ell}=1$ for all $\ell$ and that the rescaling of the fields  is trivial, i.e. $\phi^{\prime}_q=\phi_q$.
The RG equations for $\mu_{\ell}$ can be obtained by proceeding as for the case $m<d-1$ and investigating the contributions of the integral $I_{d-1}$ to the Hamiltonian. In particular, if $q$ is a type $\ell=1$ simplex, the term proportional to $\phi_q^2$ transforms as, 
\bea
\left\{\left[(1-\mu_1^{\prime})\right]\phi_q^2\right\}=\left\{\left[(1-\mu_2)+(1-\mu_1)-\frac{d}{d-\mu_1}\right]\phi_q^2\right\}.
\label{d_1A}
\eea
If instead the $(d-1)$-simplex $q$ is of type $\ell>1$, after one RG step we have,
\bea
\left\{\left[(1-\mu_{\ell}^{\prime})+(1-\mu_{\ell-1}^{\prime})\right]\phi_q^2\right\}=\left\{\left[(1-\mu_{\ell})+(1-\mu_{\ell+1})\right]\phi_q^2\right\}.
\label{d_1B}
\eea
In fact, any $(d-1)$-dimensional simplex $q$ of type $\ell>1$ after the RG step is incident exclusively to a $d$-dimensional simplex of type  $\ell$ and another $d$-dimensional simplex of type $\ell-1$.
Eqs. (\ref{d_1A}) and (\ref{d_1B}) can be solved and reduce to the single RG equation valid for $m=d-1$ 
\bea
\mu_{\ell}^{\prime}=\mu_{\ell+1}+(-1)^{\ell}\left[ (1-\mu_1)-\frac{d}{d-\mu_1}\right].
\eea

\subsection{The free-energy density and spectral dimension for $m= d-1$}

For $m=d-1$ the RG  flow is dictated by the Eqs. (\ref{RG2A}) and there is no rescaling of the fields. In this case the free-energy can be calculated using  Eq. (\ref{fg1}) with  $g(\bm{\omega})$ given by 
\bea
g(\bm{\omega})=\frac{{\mathcal N}_n}{2N_n}\ln G(\mu_1)+c,
\label{gwb}
\eea
where $c$ is a constant. Note that this expression for $g(\bm{\omega})$ differs from Eq. (\ref{gw}) as it does not contain the terms related to rescaling of the fields. 
Using this expression and the Eq. (\ref{fg1}) we can approximate the free energy $f$ by,

\bea
f&\simeq&\sum_{\tau=0}^{\infty}\frac{g(R^{(\tau)}\bm{\omega})}{d^{\tau}}= \sum_{\tau=0}^{\infty}\frac{1}{d^{\tau}}\left\{
\frac{(d-1)}{2d}\ln G\left(\mu_1^{(\tau)}\right)\right\},
\eea
with $G(\mu_1)$ given by Eq.(\ref{G_d_1}).
Using Eq. (\ref{rm0}) we can deduce the spectral density $\bar{\rho}(\mu)$ given by 
\bea
&&\hspace{-20mm}\rho(\mu)\simeq \frac{2}{\pi}\mbox{Im}\sum_{\tau=0}^{\infty}\frac{1}{d^{\tau}}\frac{\partial g(\mu_1^{(\tau)},p_2^{\tau})}{\partial \mu}\nonumber \\&&\hspace{-10mm}\simeq \frac{2}{\pi}\mbox{Im}\sum_{\tau=0}^{\infty}\frac{\lambda^{\tau}}{d^{\tau}}\frac{(d-1)}{2d}\left[{{d-1}\choose{m}}\frac{1}{d-(d-m)\mu_1^{(\tau)}}+{{d-1}\choose{m-1}}\frac{1}{\mu_1^{(\tau)}}\right].\label{rhodm1A}
\eea

\section{RG flow for the Apollonian simplicial complex}

In this  section we will investigate the RG flow for the spectrum of the $m$ dimensional up-Laplacians on a $d$-dimensional Apollonian simplicial complex and we will derive its density of eigenvalues and its spectral dimension. Interestingly, the RG  equations can be easily treated in full generality by considering the cases $m=d-1,m=d-2$ and $m\leq d-3$. 

\subsection{Case  $m=d-1$}
The RG equations for the case $m=d-1$ are given by Eq.(\ref{RG2A}), which we will repeat here for convenience
\bea
p_{\ell}=p_1=1\nonumber \\
\mu_{\ell}^{\prime}=\mu_{\ell+1}+(-1)^{\ell}\left[(1-\mu_1)-\frac{d}{d-\mu_1}\right].
\eea
The initial condition is $\mu_{\ell}=\mu\ll1$.
From these equations we can obtain the recursive equation for $\mu_1^{(\tau)}$ indicating the value of $\mu$ at the iteration $\tau$ of the RG transformation. This equation reads
\bea
\mu_{1}^{(\tau+1)}=2\mu-\mu_{1}^{(\tau)}-\left[(1-\mu_1)-\frac{d}{d-\mu_1}\right], 
\eea
where $\mu_1^{(0)}=\mu\ll 1$.
The fixed points of this RG flow  are given by  
\bea
\mu_1^{\star}&=&\frac{1}{2}\left(d-1+2\mu\right)-\frac{1}{2}\sqrt{\left(d-1+2\mu\right)^2-8d\mu}\label{fix1} \\
&=&2\frac{d}{d-1}\mu+\mathcal{O}(\mu^2),\\
\mu_1^{\star}&=&\frac{1}{2}\left(d-1+2\mu\right)+\frac{1}{2}\sqrt{\left(d-1+2\mu\right)^2-8d\mu},\label{fix2} \\
&=&d-1-\frac{2}{d-1}\mu+\mathcal{O}(\mu^2)
\eea
The relevant fixed point is defined in Eq.(\ref{fix1}), the  derivative of the recursive RG equation close to this fixed point at $\mu_1^{\star} $ is given by 
\bea
\lambda_1=\frac{d}{(d-\mu^{\star})^2}\simeq \frac{1}{d}+\frac{4}{d(d-1)}\mu+\mathcal{O}(\mu^2).
\eea
Since $\lambda_1<1$ it follows that  the fixed point $\mu_1^{\star}$ defined in Eq.(\ref{fix1}) is attractive. 
Consequently, the RG flow starting from $\mu\ll1 $ converges fast towards the fixed point $\mu_1^{\star}$ defined in Eq. (\ref{fix1}). The fixed point $\mu_1^{\star}$ is of the same order of magnitude as the initial condition for $\mu$.

In this case the fixed point is not at zero but at $\mu_1^{\star}=\mathcal{O}(\mu)$. Moreover, the fixed point is attractive. This constitute a rather special scenario that we will not find for smaller values of $m$. A careful study of the equation $(\ref{rhodm1A})$ for the spectral density $\rho(\mu)$  reveals that in this case  the corresponding up-Laplacian does not display a finite spectral dimension.

\subsection{Case  $m=d-2$}

\begin{figure}[h!]
	\begin{center}
 \includegraphics[width=0.97\columnwidth]{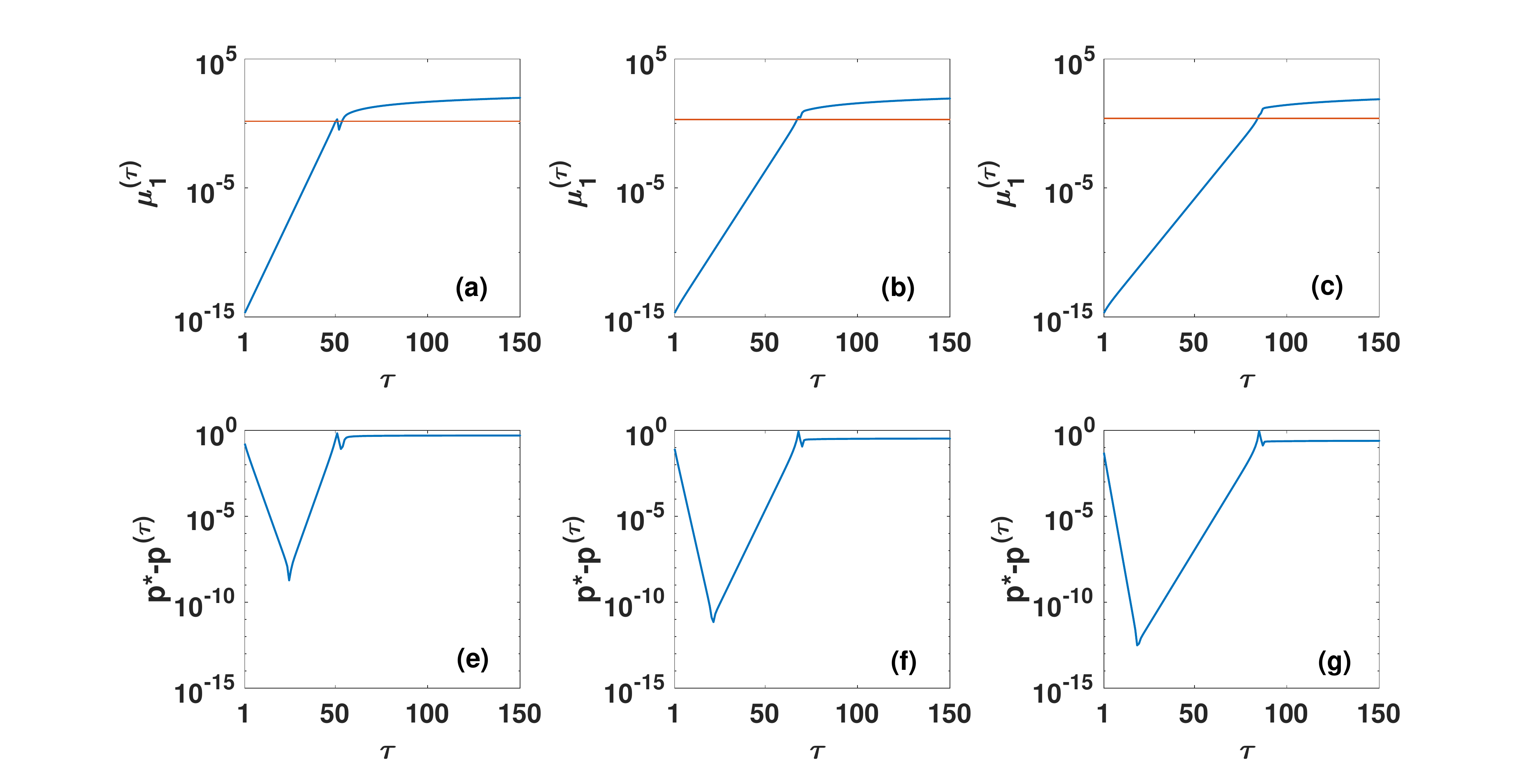}	
  \end{center}\
  \caption{The RG flow for the Apollonian simplicial complex for $m=d-2$ is represented by plotting the numerically integrated values of $\mu_1^{(\tau)}$ and $p^{\star}-p^{\tau}$ versus $\tau$, where $p^{\star}=(d-1)/d$ (blue curves). The red curve indicates the constant value $\mu=d/2$, where Eq.(\ref{bar_rhoe}) has a simple pole. Plots (a)-(e), (b)-(f) and (c)-(g) display the RG flow for dimension $d=2$, $d=3$ and $d=4$ respectively. In all the plots $\mu=10^{-15}$.}
  	\label{fig:RGflow}
	\end{figure}

For $m=d-2$ the RG Eqs. (\ref{RGhdA}) imply that 
\bea 
p_{\ell}&=&p\nonumber\\
\mu_{\ell}&=&\mu_2
\eea
for all $\ell\geq 1$, while  $\mu_1$ and $p$ obey the following recursive RG equations, 
\bea
&&(1-\mu_{1}')=\left((1-\mu_1)+(1-\mu_{2})p-\frac{d-1}{d-2\mu_1}\right) \left[p+\frac{1}{d-2\mu_1}\right]^{-1},\nonumber\\
&&\mu_2^{\prime}=\mu_2,\nonumber\\
&&p^{\prime}=p\left[p+\frac{1}{d-2\mu_1}\right]^{-1},
\label{RGd2}
\eea
with initial condition  $(\mu_{\ell},p_{\ell})=(\mu,1)$ with $\mu\ll 1$  for all $\ell\geq 1$.
In the zero order approximation we can put $\mu_2=\mu=0$. Therefore   the renormalization group equations (\ref{RGd2}) have three fixed points:
\bea
(\mu^{\star},p^{\star})&=&(0,0),\label{1f} \\
(\mu^{\star},p^{\star})&=&\left(0,\frac{d-1}{d}\right),\label{2f} \\
(\mu^{\star},p^{\star})&=&\left(\frac{d+1}{2},0\right).\label{3f} 
\eea
Close to the fixed point defined in Eq.(\ref{2f}) the linearised RG equations read,
\bea
\hspace{-15mm}\left(\begin{array}{c}\mu_1^{\prime}\\p^{\prime}-\frac{d-1}{d}\end{array}\right)=
\left(\begin{array}{cc}(2+d)/d&0\\-2(d-1)/d^3&1/d\end{array}\right)\left(\begin{array}{c}\mu_1\\p-\frac{d-1}{d}\end{array}\right)+\mu\frac{d-1}{d}\left(\begin{array}{c}1\\0\end{array}\right).
\eea
It follows that the  eigenvalues of the Jacobian are 
\bea
\lambda_1=\lambda=1+\frac{2}{d},\\
\lambda_2=\frac{1}{d},
\eea
i.e. close to the fixed point defined in Eq. (\ref{2f}) there is one attractive and one repulsive direction.
For initial conditions $(\mu,p)=(\mu,1)$, with $\mu\ll 1$, the RG flow approaches the  fixed point defined in Eq.(\ref{2f}) and then runs away following the repulsive direction towards the fixed point defined in Eq.(\ref{3f}).
Since  $\mu^{\star}$ at the  fixed  {point} defined in Eq.(\ref{3f}) is close to the pole of Eq.(\ref{bar_rhoe}) determining the asymptotic scaling of $\rho(\mu)$, the RG flow close to the pole $\mu_1^{(\tau)}\simeq d/(d-m)=d/2$  cannot be approximated by scaling  Eq. (\ref{gflow1}) determined by the second fixed point (defined in Eq. (\ref{2f})).
This scenario can be deduced by the direct numerical implementation of the RG flow shown in Figure \ref{fig:RGflow}, where plot $\mu_1^{(\tau)}$, and $p^{(\tau)}$ versus $\tau$, for different dimensions $d=2,3,4$.
From the plots of $p^{\star}-p^{(\tau)}$ versus $\tau$ where $p^{\star}=(d-1)/d$  we observe the initial approach  of the RG flow to the fixed point defined in Eq.(\ref{2f}) and the  subsequent repulsion of the RG flow away from it as $p^{\star}-p$ first decreases exponentially then increases exponentially with $\tau$. Moreover, from the plots showing $\mu_1^{(\tau)}$ versus $\tau$, it is clear to see that as $\mu_1^{(\tau)}$ approaches the pole of Eq.(\ref{bar_rhoe}), i.e. $\mu^{(\tau)}=d/2$ (red line), the RG flow deviates from the exponential growth and starts to be affected by the fixed point defined in Eq.(\ref{3f}).

Using Eq. (\ref{ds0}) one would expect the spectral dimension $d_S$ is given by 
\bea
d_S=2\frac{\ln d}{\ln \lambda}=2\frac{\ln d}{\ln [1+2/d]}.
\eea
However, this is incorrect, because the RG flow is affected by  {the fixed point defined in Eq. (\ref{3f})} close to the pole at $\mu_1^{(\tau)}\simeq d/(d-m)=d/2$ of the explicit expression for $\bar{\rho}(\mu)$ in Eq. (\ref{bar_rhoe}). 
A detailed prediction of the spectral dimension could be eventually predicted by studying the RG flow numerically, this type of investigation is left for subsequent studies.

\subsection{Case  $m\leq d-3$}

For deriving the RG flow for the case $m\leq d-3$ we can rewrite the RG Eqs.(\ref{RGhdA}) in a simplified way with,
\bea
x_{\ell}=(1-\mu_{\ell})p_{\ell}.
\label{xell}
\eea
We obtain a new set of RG equations relating the parameters   $(\{x_{\ell}^{(\tau)}\},\{p_{\ell}^{(\tau)}\})$ at iteration $\tau$  of the RG transformation with the parameters $(\{x_{\ell}^{(\tau+1)}\},\{p_{\ell}^{(\tau+1)}\})$ at the next RG iteration. This set of equations is given by 
\bea
&&\hspace{-25mm}x_{\ell}^{(\tau+1)}=\left[x_{\ell}^{(\tau)}+\left(x_1^{(\tau)}-\frac{(m+1)}{m+(d-m)x_1^{(\tau)}}\right)(d-m-2)^{\ell-1}\right]\left[p_2^{(\tau)}+\frac{d-m-1}{m+(d-m)x_1^{(\tau)}}\right]^{-1},\nonumber \\
&&\hspace{-25mm}p_{\ell}^{(\tau+1)}=\left[p_{\ell+1}^{(\tau)}+\frac{(d-m-1)(d-m-2)^{\ell-1}}{m+(d-m)x_1^{(\tau)}}\right]\left[p_2^{(\tau)}+\frac{d-m-1}{m+(d-m)x_1^{(\tau)}}\right]^{-1},
\label{RGhd3}
\eea
with initial conditions $x_{\ell}^{(0)}=1-\mu$ and $p_{\ell}^{(0)}=1$.
In order to find the solution of these equations we use the  auxiliary variable $y_1^{(\tau)}$ given by  
\bea
{y}_{1}^{(\tau+1)}=p_{2}^{(\tau+1)}+\frac{d-m-1}{m+(d-m)x_1^{(\tau+1)}}.
\eea 
The explicit solution of the RG equations (\ref{RGhd3})   reads
\bea
&&\hspace{-25mm}p_{2}^{(\tau+1)}=\prod_{m=1}^{\tau}\frac{1}{y_1^{(m)}}+{(d-m-1)}\sum_{m=1}^{\tau}\frac{(d-m-2)^{\tau-m+1}}{m+(d-m)x^{(m)}}\prod_{m'=m}^{\tau}\frac{1}{y_1^{(m')}},\nonumber\\ 
&&\hspace{-25mm}{y}_{1}^{(\tau+1)}=p_{2}^{(\tau+1)}+\frac{d-m-1}{m+(d-m)x_1^{(\tau+1)}},\nonumber\\
&&\hspace{-15mm}=\prod_{m=1}^{\tau}\frac{1}{y_1^{(m)}}+{(d-m-1)}\sum_{m=1}^{\tau}\frac{(d-m-2)^{\tau-m+1}}{m+(d-m)x^{(m)}}\prod_{m'=m}^{\tau}\frac{1}{y_1^{(m')}}+\frac{d-m-1}{m+(d-m)x_1^{(\tau+1)}},\nonumber \\
&&\hspace{-25mm}x_{1}^{(\tau+1)}=x_1^{(1)}\prod_{m=1}^{\tau}\frac{1}{y_1^{(m)}}+\sum_{m=1}^{\tau}\left(x_1^{(m)}-\frac{(m+1)}{m+(d-m)x_1^{(m)}}\right)(d-m-2)^{\tau-m}\prod_{m'=m}^{\tau}\frac{1}{y_1^{(m')}}.\nonumber 
\eea
This solution shows that $p_2^{(\tau+1)},y_1^{(\tau+1)}$ and $x_1^{(\tau+1)}$ depend on the entire RG flow up to time $\tau$, i.e. on all the values of the parameters $p_2^{(\tau')},y_1^{(\tau')}$ and $x_1^{(\tau')}$ with $\tau'\leq \tau$. This solution therefore seems to indicate that in order to calculate the value of $x_{\ell}^{(\tau+1)}$ and $p_{\ell}^{(\tau+1)}$ the knowledge of the entire RG flow up to iteration $\tau$ is necessary. However, one can recover some Markovian recursive equations by introducing  the additional auxiliary variables called $A^{(\tau)},B^{(\tau)}$ and $C^{(\tau)}$. 
The  auxiliary variables $A^{(\tau)},B^{(\tau)}$ and $C^{(\tau)}$ are defined as
\bea
A^{(\tau)}&=&\prod_{m=1}^{\tau}\frac{1}{y_1^{(m)}},\nonumber \\
B^{(\tau)}&=&{(d-m-1)}\sum_{m=1}^{\tau}\frac{(d-m-2)^{\tau-m+1}}{m+(d-m)x^{(m)}}\prod_{m'=m}^{\tau}\frac{1}{y_1^{(m')}}+\frac{d-m-1}{m+(d-m)x_1^{(\tau+1)}},
\nonumber \\C^{(\tau)}&=&\sum_{m=1}^{\tau}\left(x_1^{(m)}-\frac{(m+1)}{m+(d-m)x_1^{(m)}}\right)(d-m-2)^{\tau-m}\prod_{m'=m}^{\tau}\frac{1}{y_1^{(m')}}.\nonumber 
\eea
The variables $y_1^{({\tau+1)}}$ and $x_{1}^{(\tau+1)}$ can be simply expressed in terms of $A^{(\tau)},B^{(\tau)}$ and $C^{(\tau)}$ by
\bea
y_1^{(\tau+1)}&=&A^{(\tau)}+B^{(\tau)},\nonumber\\
x_{1}^{(\tau+1)}&=&(1-\mu)A^{(\tau)}+C^{(\tau)}.
\label{x1y1Ap}
\eea
The solution of the RG equations can be written as the following set of   recursive equations for $A^{(\tau)},B^{(\tau)}$ and $C^{(\tau)}$ 
\bea
A^{(\tau+1)}&=&\frac{1}{y_1^{(\tau+1)}}A^{(\tau)}
,
\nonumber \\
B^{(\tau+1)}&=&\frac{d-m-2}{y_1^{(\tau+1)}}B^{(\tau)}+({d-m-1})\frac{1}{m+(d-m)x_1^{(\tau+2)}}
,
\nonumber \\
C^{(\tau+1)}&=&\frac{(d-m-2)}{y_1^{(\tau+1)}}C^{(\tau)}+\frac{1}{y_1^{(\tau+1)}}\left(x_1^{(\tau+1)}-\frac{(m+1)}{m+(d-m)x_1^{(\tau+1)}}\right)
.
\label{127}
\eea
This set of equations can be written as a closed set of equations for $A^{(\tau)},B^{(\tau)}$ and $C^{(\tau)}$ using Eq. (\ref{x1y1Ap}), 
with initial conditions $A^{(0)}=1,B^{(0)}=(d-m-1)/(d-(d-m)\mu)),C^{(0)}=0$.\\

The fixed point of these RG equations at $\mu=0$ is 
\bea
A^{\star}=0
,
\nonumber \\
B^{\star}=\frac{d^2-(m+1)(d+1)}{d}
,
\nonumber \\
C^{\star}=1
.
\eea
The Jacobian matrix of these RG equations has  eigenvalues $\lambda_1>\lambda_2>\lambda_3$ given by 
\bea
\lambda_1=\lambda=\frac{d^2-m(d+1)}{d^2-(m+1)(d+1)}
,
\nonumber \\
\lambda_2=\frac{d}{d^2-(m+1)(d+1)}
,
\nonumber \\
\lambda_3=0
\label{lmld3}
\eea
with $\lambda_1>1$ and $\lambda_2<1$.

The right eigenvectors corresponding to these eigenvalues are
\bea
{\bf u_1}&=&\frac{1}{c_1}\left(d^2+d-m(1-d+m),-d^2,d^3-d^2+d-m(1-d+d^2+m)\right),\nonumber \\
{\bf u_2}&=&\left(1,0,0\right),\nonumber \\
{\bf u_3}&=&\frac{1}{c_3}\left(d^3-d^2-d+m(1-2d-d^2+m),d^2,d^2-d+m(1-2d+m)\right),\nonumber 
\eea
where $c_1$ and $c_3$ are normalization constants.
The left eigenvectors corresponding to these eigenvalues are
\bea
{\bf v_1}&=&\frac{1}{d_1}\left(0,d^2-d+m(1-2d+m),-d^2\right),\nonumber \\
{\bf v_2}&=&\frac{1}{d_2}\left(-1,d-2-m,1\right),\nonumber \\
{\bf v_3}&=&\frac{1}{d_3}\left(0,d^3-d^2+d-m(1-d+d^2+m),d^2\right),\nonumber 
\eea
where $d_1,d_2,d_3$ are normalization constants.
In order to solve Eqs.(\ref{127}) we indicate with   ${\bf X}^{(\tau)}$ the column vector 
\bea
{\bf X}^{(\tau)}=\left(A^{(\tau)},B^{(\tau)},C^{(\tau)}\right)
.
\eea
By linearizing Eqs.(\ref{127}) near the fixed point ${\bf X}^{\star}$ given by 
\bea
{\bf X}^{\star}=\left(A^{\star},B^{\star},C^{\star}\right)
,
\eea
we obtain
\bea
{\bf X}^{(\tau)}={\bf X}^{\star}+\sum_{m=1}^3\lambda_m^{\tau}{\bf v}_m\avg{{\bf u}_m,{\bf X}^{(0)}-{\bf X}^{\star}}.
\eea
For the  leading order term, we have 
\bea
{\bf X}^{(\tau)}={\bf X}^{\star}+\lambda_1^{\tau}{\bf v}_1\avg{{\bf u}_1,{\bf X}^{(0)}-{\bf X}^{\star}}
,
\eea
where the scalar product is,
\bea
\avg{{\bf u}_1,{\bf X}^{(0)}-{\bf X}^{\star}}\propto \frac{\mu}{d-\mu(d-m)}.
\eea
We therefore have proved that for $\mu\ll 1$ we have, 
\bea
\mu_1^{(\tau)}\propto \lambda^{\tau}\mu.
\eea
Using Eq. (\ref{ds0}) it follows that for $m\leq d-3$  the spectral dimension $d_S$ decreases with increasing $m$ and is given by 
\bea
d_S=2\frac{\ln d}{\ln \lambda}=2(\ln d) \left[\ln\left(\frac{d^2-m(d+1)}{d^2-(m+1)(d+1)}\right)\right]^{-1}.
\eea
Finally, we observe that in the limit $d\to \infty$ and $m\ll d$  the spectral dimension  scales like
\bea
d_S\simeq (2\ln d)\left[d-m-\frac{3}{2}+O(1/d)\right].
\eea
The spectral dimension therefore grows faster than linearly with the topological dimension $d$.

\section{ General RG equations for  the pseudo-fractal simplicial complex}
\subsection{The RG equations}

In a $d$-dimensional  pseudo-fractal simplicial complex at each iteration $n$ each $(d-1)$-simplex  is glued to a new $d$-dimensional simplex. The difference with the algorithm generating the Apollonian simplicial complexes is that in the case of the Apollonian simplicial complex  at each iteration $n$ only the $(d-1)$-simplices of the last generation are glued to a new $d$-dimensional simplex.
Given the structure of the pseudo-fractal simplicial complex and its relation to the Apollonian simplicial complex, which was already noted in Ref.\cite{RG1}, the general  RG equations for  the pseudo-fractal simplicial complex can be easily derived from those for the Apollonian simplicial complex. In fact it is sufficient to observe that in  the pseudo-fractal simplicial complex each simplex of type $\ell$ receives the sum of the contributions coming from the integration of the Gaussian variables associated to the $d$-simplices added at the last generation.
The  RG equations for $m\leq d-2$ are therefore given by 
\bea
&&\hspace{-20mm}(1-\mu_{\ell}')p_{\ell}^{\prime}=\left[(1-\mu_{\ell+1})p_{\ell+1}+\left((1-\mu_1)-\frac{(m+1)}{d-(d-m)\mu_1}\right)\sum_{\ell^{\prime}=0}^{\ell-1}(d-m-2)^{\ell^{\prime}}\right]\nonumber \\
&&\times \left[p_2+\frac{(d-m-1)}{d-(d-m)\mu_1)}\right]^{-1},\nonumber \\
&&\hspace{-20mm}p'_{\ell}=\left[p_{\ell+1}+\frac{(d-m-1)}{d-(d-m)\mu_1}\sum_{\ell^{\prime}=0}^{\ell-1}(d-m-2)^{\ell^{\prime}}\right]\left[p_2+\frac{(d-m-1)}{d-(d-m)\mu_1}\right]^{-1},
\label{RGps}
\eea
for all $\ell\geq 1$, with initial conditions  $(\mu_{\ell},p_{\ell})=(\mu,1)$ with $\mu\ll 1$ for all $\ell\geq 1$.
For $m=d-1$ every $m$-simplex of type $\ell$ is connected to a $d$-simplex of generation $n$ and the RG equations for $m=d-1$ and $\ell\geq 1$ read 
\bea
p_{\ell}=p_1=1 \eea
and 
\bea
(1-\mu_{\ell}^{\prime})=(1-\mu_{\ell+1})+\left[(1-\mu_1)-\frac{d}{d-\mu_1}\right]
\label{RG1ps1}
\eea
with initial conditions $(\mu_{\ell},p_{\ell})=(\mu,1)$ with $\mu\ll 1$ for all $\ell\geq 1$.

\subsection{The free-energy density and the spectral dimension}

The free energy is given by Eq.  (\ref{fg2}). By using  a procedure similar to the one used to derive  the corresponding expression for the Apollonian simplicial complex we  easily find  for $m\leq d-2$
\bea
g(\bm{\omega})=\frac{{\mathcal N}_n}{2N_n}\ln G(\mu_1)+\frac{N_{n-1}}{2N_{n}}\ln \left[p_2+\frac{d-m-1}{d-(d-m)\mu_1)}\right]+c,
\label{gw2}
\eea
where $c$ is a constant.
Given this expression for $g(\bm{\omega})$, the free energy density $f$
 obtained from Eq. (\ref{fg2}) reads
\bea
\hspace{-20mm}f\simeq\sum_{\tau=0}^{\infty}\frac{g(R^{(\tau)}\bm{\omega})}{(d+1)^{\tau}}\nonumber \\ &&\hspace{-20mm}\simeq \sum_{\tau=0}^{\infty}\frac{1}{(d+1)^{\tau}}\left\{
\frac{d}{2(d+1)}\ln G\left(\mu_1^{(\tau)}\right)+\frac{1}{2(d+1)}\ln \left[p_2^{(\tau)}+\frac{d-m-1}{d-(d-m)\mu_1^{(\tau)}}\right]\right\}.\nonumber
\eea
For the pseudo-fractal complex, we expect to find a  relevant repulsive fixed point at $(\mu^{\star},p_2^{\star})=(0,p^{\star})$. Under this hypothesis the RG flow is described by 
\bea
\mu_1^{(\tau)}\simeq\mu\lambda^{\tau}\nonumber \\
p_2^{(\tau)}\simeq p^{\star}+\lambda^{\tau}(1-p^{\star})
\label{gflow2}
\eea
close to the relevant  fixed point, where $\lambda>1$ is the largest eigenvalue of the linearized RG equations close to the relevant fixed point.
Using Eq. (\ref{rm0}), the spectral density $\bar{\rho}(\mu)$ can be expressed as
\bea
&&\hspace{-27mm}\bar{\rho}(\mu)\simeq \frac{2}{\pi}\mbox{Im}\sum_{\tau=0}^{\infty}\frac{1}{(d+1)^{\tau}}\frac{\partial g(\mu_1^{(\tau)},p_2^{\tau})}{\partial \mu}\nonumber \\&&\hspace{-27mm}\simeq \frac{2}{\pi}\mbox{Im}\sum_{\tau=0}^{\infty}\frac{\lambda^{\tau}}{(d+1)^{\tau}}\frac{d}{2(d+1)}\left[{{d-1}\choose{m}}\frac{1}{d-(d-m)\mu_1^{(\tau)}}+{{d-1}\choose{m-1}}\frac{1}{\mu_1^{(\tau)}}\right]\nonumber \\
&&\hspace{-27mm}+ \frac{2}{\pi}\mbox{Im}\sum_{\tau=0}^{\infty}\frac{\lambda^{\tau}}{(d+1)^{\tau}}\frac{d-m}{2(d+1)}y\left[\left(p_2^{(\tau)}\left[d-(d-m)\mu_1^{(\tau)}\right]+y\right)\left(d-(d-m)\mu_1^{(\tau)}\right)\right]^{-1},
\eea
where $y=d-m-1$.
In the pseudo-fractal simplicial complex, the spectrum of the up-Laplacian of order $m$ acquires a delta peak at $\mu=0$ as well. This corresponds to the finite density of zero eigenvalues of the up-Laplacian, i.e.
\bea
\bar{\rho}(\mu)=\bar{\rho}(0)\hat{\delta}(\mu)+{\rho}(\mu)
\label{rho_delta}
\eea
where  $\bar{\rho}(0)$ given by 
\bea
\bar{\rho}(0)=\frac{d}{d+1}{{d-1}\choose{m-1}}\frac{1}{1-1/(d+1)}.
\eea
and  the regular part of the spectrum is given by 
\bea
&&\hspace{-25mm}{\rho}(\mu)\simeq \frac{2}{\pi}\mbox{Im}\sum_{\tau=0}^{\infty}\frac{\lambda^{\tau}}{(d+1)^{\tau}}\frac{d}{2(d+1)}\left[{{d-1}\choose{m}}\frac{1}{d-(d-m)\mu_1^{(\tau)}}\right]\nonumber \\
&&\hspace{-28mm}+ \frac{2}{\pi}\mbox{Im}\sum_{\tau=0}^{\infty}\frac{\lambda^{\tau}}{(d+1)^{\tau}}\frac{d-m}{2(d+1)}y\left[\left(p_2^{(\tau)}\left[d-(d-m)\mu_1^{(\tau)}\right]+y\right)\left(d-(d-m)\mu_1^{(\tau)}\right)\right]^{-1}.
\label{ploe}
\eea
By approximating this expression with an integral over   $\tau$ and by changing the variable of this integral to   $z=\lambda^{\tau}$ we can approximate $\rho(\mu)$ by using the residue theorem at the pole $\mu_1^{(\tau)}=z\mu=d/(d-m)$, obtaining  the asymptotic scaling 
\bea
{\rho}(\mu)&\simeq& C \mu^{d_S/2-1}.
\eea
The spectral dimension $d_S$ is then given by  
\bea
d_S=2\frac{\ln (d+1)}{\ln \lambda}.
\label{ds0ps}
\eea
For $m=d-1$ the Gaussian fields are not rescaled and $g(\bm{\omega})$ is given by 
\bea
g(\bm{\omega})=\frac{{\mathcal N}_n}{2N_n}\ln G(\mu_1)+c,
\label{gwb}
\eea
where $c$ is a constant. Using this expression and Eq. (\ref{fg2}) we can approximate the free energy $f$ by

\bea
f&\simeq&\sum_{\tau=0}^{\infty}\frac{g(R^{(\tau)}\bm{\omega})}{(d+1)^{\tau}}= \sum_{\tau=0}^{\infty}\frac{1}{(d+1)^{\tau}}\left\{
\frac{d}{2(d+1)}\ln G\left(\mu_1^{(\tau)}\right)\right\},
\eea
with $G(\mu_1)$ given by Eq.(\ref{G_d_1}).
Using Eq. (\ref{rm0}), we can deduce that the spectral density $\bar{\rho}(\mu)$ is given by 
\bea
&&\hspace{-20mm}\rho(\mu)\simeq \frac{2}{\pi}\mbox{Im}\sum_{\tau=0}^{\infty}\frac{1}{(d+1)^{\tau}}\frac{\partial g(\mu_1^{(\tau)},p_2^{\tau})}{\partial \mu}\nonumber \\&&\hspace{-20mm}\simeq \frac{2}{\pi}\mbox{Im}\sum_{\tau=0}^{\infty}\frac{\lambda^{\tau}}{(d+1)^{\tau}}\frac{d}{2(d+1)}\left[{{d-1}\choose{m}}\frac{1}{d-(d-m)\mu_1^{(\tau)}}+{{d-1}\choose{m-1}}\frac{1}{\mu_1^{(\tau)}}\right].\label{polem1} 
\eea
\section{RG flow for the pseudo-fractal simplicial complex}
In this section we will treat the RG flow for the pseudo-fractal simplicial complex.
We consider  the cases $m=d-1,m=d-2,m=d-3$ and $m<d-3$.

  \subsection{Case $m=d-1$}
\begin{figure}[h!]
	\begin{center}
 \includegraphics[width=0.97\columnwidth]{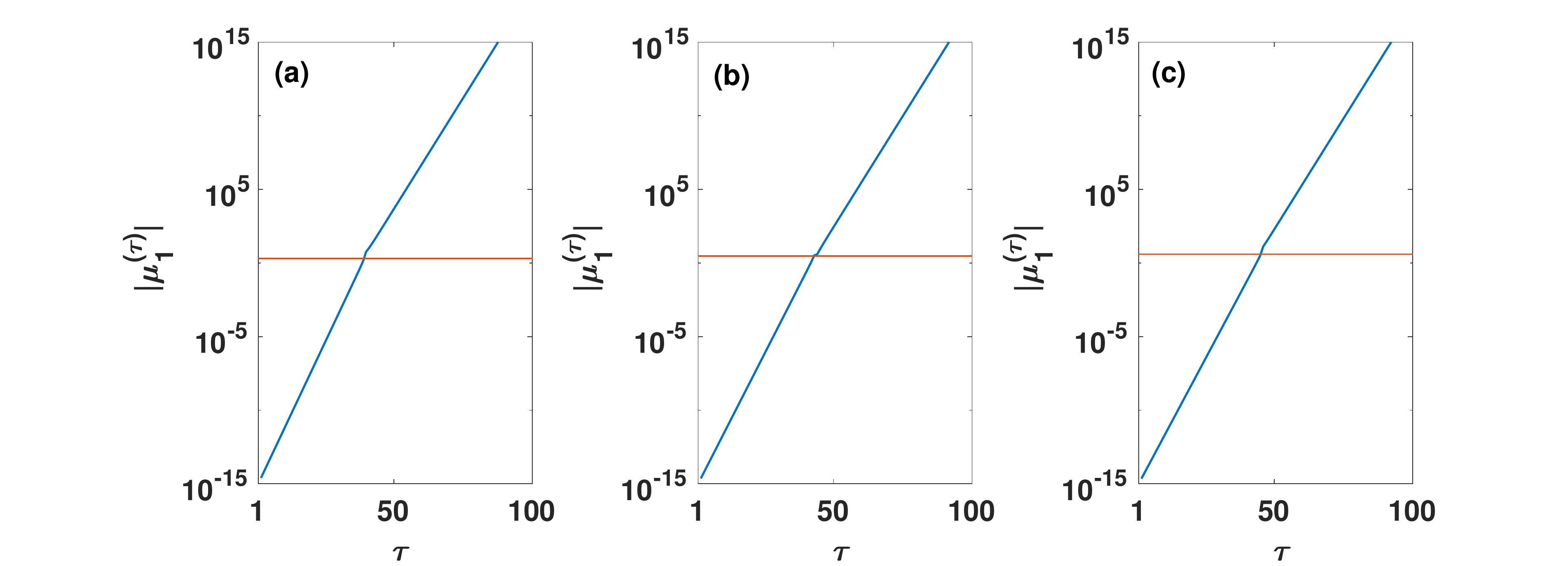}	
  \end{center}\
  \caption{The RG flow of the pseudo-fractal simplicial complex for $m=d-1$ is represented by plotting the numerically integrated values of $|\mu_1^{(\tau)}|$  versus $\tau$ (blue curves). The red curves indicate the constant value $\mu_1=d$ where Eq.(\ref{polem1}) has a simple pole. Plots (a), (b) and (c) display RG flow for dimension $d=2$, $d=3$ and $d=4$ respectively. In all the plots $\mu=10^{-15}$.}
  	\label{fig:RGflowps}
	\end{figure}

  The RG equations for $m=d-1$ are given by Eqs.(\ref{RG1ps1}), which can be used to derive the following recursive equation for $\mu_1$,
\bea
\mu_1^{\prime}&=&\mu_1-\left[(1-\mu_1)-\frac{d}{d-\mu_1}\right],
\label{mud1s}
\eea
with initial condition $\mu_1=\mu.$
The  fixed points  of this equations are  
  \bea
  \mu_1^{\star}=0,\label{fix1ps} \\
  \mu_1^{\star}=d+1.\label{fix2ps}
  \eea
At the fixed point at $\mu^{\star}=0$ the recursive equation Eqs.(\ref{mud1s}) has eigenvalue
\bea\lambda=2+\frac{1}{d}>1,
\eea
so $\mu^{\star}=0$ is a repulsive fixed point.
The RG flow starts from $\mu_1=\mu\ll1$ and runs away from $\mu^{\star}=0$ according to
\bea
\mu^{(\tau)}=\mu \lambda^{\tau}.
\label{rgex}
\eea
In doing so,  {the RG flow approaches the singularity of Eq.(\ref{mud1s}) at $\mu_1=d$ and the linearised RG flow described by Eq. (\ref{rgex}) is not longer valid. Therefore the RG flow  changes its trend, in some cases even changing sign.} This scenario is apparent from Figure $\ref{fig:RGflowps}$ were the absolute values of $\mu_1^{(\tau)}$ (indicating the value of the parameter $\mu_1$ at iteration $\tau$ of the RG transformation) are plotted versus $\tau$
This is a situation analogous to the case $m=d-2$ for the Apollonian simplicial complexes, where the RG flow changes trends very close to the pole in Eq.(\ref{polem1}). In this case $\lambda$ cannot be used to give a good estimation of the spectral dimension $d_S$.

\subsection{Case $m=d-2$}

For $m=d-2$ the RG Eqs.(\ref{RGps}) for the pseudo-fractal simplicial complex greatly simplify.  We have 
\bea
\mu_{\ell}=\mu_1
\eea
for $\ell\geq 1$ and 
\bea
p_{\ell}=p\nonumber \\
\eea
for all $\ell\geq 2$.
The resulting RG equations are
\bea
&&(1-\mu_1')=\left[2(1-\mu_1)-\frac{d-1}{d-2\mu_1}\right]\left[1+\frac{1}{d-2\mu_1}\right]^{-1},\nonumber\\
&&p'=1,
\label{PSRGd2}
\eea
with initial conditions $(\mu_{\ell},p_{\ell})=(\mu,1)$ with $\mu\ll 1$ for all $\ell\geq 1$.
The fixed point is $(\mu^{\star},p^{\star})=(0,1)$.
The eigenvalue of this system of equations is 
\bea
\lambda=2.
\eea 
The fixed point is $\mu^{\star}_1=0$ and $p^{\star}=1$ with eigenvalue $\lambda=2$. 
Using Eq.(\ref{ds0ps}) we can predict the spectral dimension
\bea
d_S=2\frac{\ln (d+1)}{\ln \lambda}=2\frac{\ln (d+1)}{\ln 2}.
\eea

\subsection{Case $m=d-3$}

In the case $m=d-3$ the RG Eqs.(\ref{RGps}) can be expressed in terms of the variables $x_{\ell}$ as defined in Eq. (\ref{xell}). Using $x_{\ell}^{(\tau)}$ and $p_{\ell}^{(\tau)}$ for indicating the parameter values at iteration $\tau$, by performing the sum over $\ell'$, the Eqs. (\ref{RGps}) for $m=d-3$ can be written as 
\bea
&&x_{\ell}^{(\tau+1)}=\left[x_{\ell+1}^{(\tau)}+\left(x_1^{(\tau)}-\frac{d-2}{m+3x_1^{(\tau)}}\right)\ell\right]\left[p_2^{(\tau)}+\frac{d-m-1}{m+3x_1^{(\tau)}}\right]^{-1},\nonumber \\
&&p_{\ell}^{(\tau+1)}=\left[p_{\ell+1}^{(\tau)}+\frac{(d-m-1)\ell}{m+3x_1^{(\tau)}}\right]\left[p_2^{(\tau)}+\frac{d-m-1}{m+3 x_1^{(\tau)}}\right]^{-1},
\label{PFhd3}
\eea
with initial conditions  $x_{\ell}^{(0)}=1-\mu$ and $p_{\ell}^{(0)}=1$.
Equations (\ref{PFhd3}) can be solved in terms of the auxiliary variable
\bea
y_1^{(\tau+1)}&=&p_2^{(\tau+1)}+\frac{2}{m+3x_1^{(\tau+1)}}
,
\eea
and we obtain
\bea
&&\hspace*{-25mm}x_1^{(\tau+1)}=(1-\mu)\prod_{m=1}^{\tau}\frac{1}{y_1^{(m)}}+\sum_{m=1}^{\tau}\left(x_1^{(m)}-\frac{d-2}{m+3x_1^{(m)}}\right)(\tau+1-m)\prod_{m'=m}^{\tau}\frac{1}{y^{(m')}}
,
\nonumber \\
&&\hspace*{-25mm}p^{(\tau+1)}_2=\prod_{m=1}^{\tau}\frac{1}{y_1^{(m)}}+({d-m-1})\sum_{m=1}^{\tau}\frac{1}{m+3x_1^{(m)}}(\tau+2-m)\prod_{m'=m}^{\tau}\frac{1}{y^{(m')}}.
\eea
Also in the pseudo-fractal case these non-Markovian equations can be turned to Markovian iterative relations by expressing the variable at iteration $\tau+1$ exclusively in terms of the variable at iteration $\tau$. This is achieved by introducing the auxiliary variables $A^{(\tau)},B^{(\tau)},C^{(\tau)},D^{(\tau)}$ and $E^{(\tau)}$ defined as
\bea
A^{(\tau)}&=&\prod_{m=1}^{\tau}\frac{1}{y_1^{(m)}},\nonumber \\
B^{(\tau)}&=&\sum_{m=1}^{\tau}\frac{2}{d+3x_1^{(m)}}(\tau+2-m)\prod_{m'=m}^{\tau}\frac{1}{y^{(m')}},\nonumber\\
C^{(\tau)}&=&\sum_{m=1}^{\tau}\left(x_1^{(m)}-\frac{d-2}{m+3x_1^{(m)}}\right)(\tau+1-m)\prod_{m'=m}^{\tau}\frac{1}{y^{(m')}},\nonumber \\
D^{(\tau)}&=&\sum_{m=1}^{\tau}\left(x_1^{(m)}-\frac{d-2}{m+3x_1^{(m)}}\right)\prod_{m'=m}^{\tau}\frac{1}{y^{(m')}},\nonumber \\
E^{(\tau)}&=&\sum_{m=1}^{\tau}\frac{2}{m+3x_1^{(m)}}\prod_{m'=m}^{\tau}\frac{1}{y^{(m')}}.
\label{def1Ad3}
\eea
These auxiliary variables are related to $y_1^{(\tau)}$ and $x_1^{(\tau)}$ by
\bea
y_1^{(\tau+1)}&=&A^{(\tau)}+B^{(\tau)}+\frac{2}{m+3x_1^{(\tau+1)}}
,
\nonumber \\
x_1^{(\tau+1)}&=&(1-\mu)A^{(\tau)}+C^{(\tau)}.
\label{158}
\eea

The recursive Markovian RG equations for the case $m=d-3$ read 
\bea
x_1^{(\tau+1)}&=&(1-\mu)A^{(\tau)}+C^{(\tau)}
,
\nonumber \\
y_1^{(\tau+1)}&=&A^{(\tau)}+B^{(\tau)}+\frac{2}{m+3[(1-\mu)A^{(\tau)}+C^{(\tau)}]}
,
\nonumber \\
A^{(\tau+1)}&=&\frac{1}{y_1^{(\tau+1)}}A^{(\tau)}
,
\nonumber \\
B^{(\tau+1)}&=&\frac{1}{y_1^{(\tau+1)}}\left[B^{(\tau)}+E^{(\tau)}+\frac{4}{m+3x_1^{(\tau+1)}}\right]
,
\nonumber \\
C^{(\tau+1)}&=&\frac{1}{y_1^{(\tau+1)}}\left[C^{(\tau)}+D^{(\tau)}+\left(x_1^{(\tau+1)}-\frac{d-2}{m+3x_1^{(\tau+1)}}\right)\right]
,
\nonumber \\
D^{(\tau+1)}&=&\frac{1}{y_1^{(\tau+1)}}\left[D^{(\tau)}+\left(x_1^{(\tau+1)}-\frac{d-2}{m+3x_1^{(\tau+1)}}\right)\right]
,
\nonumber \\
E^{(\tau+1)}&=&\frac{1}{y_1^{(\tau+1)}}\left[E^{(\tau)}+\frac{2}{m+3x_1^{(\tau+1)}}\right],
\label{RGd3psA}
\eea
with initial conditions 
$A^{(1)},B^{(1)},C^{(1)},D^{(1)}$ and $E^{(1)}$, which can be found by inserting $x_1^{(0)}=1-\mu$ and $y_1^{(0)}=1+\frac{2}{m+3(1-\mu)}$ in Eqs. (\ref{def1Ad3}) and (\ref{158}).\\

The relevant fixed point of these equations is 
\bea
A^{\star}&=&0
,
\nonumber \\
B^{\star}&=&\frac{1}{d}\left(d-1+\sqrt{1+2d}\right)
,
\nonumber \\
C^{\star}&=&1
,
\nonumber \\
D^{\star}&=&\frac{1}{d}\left(-1+{\sqrt{1+2d}}\right)
,
\nonumber \\
E^{\star}&=&\frac{1}{d}\left(-1+{\sqrt{1+2d}}\right)
.
\nonumber \\
\eea
Close to this fixed point, the RG equations (\ref{RGd3psA}) have the relevant eigenvalue
\bea
\lambda=[1+d+\sqrt{1+2d}]^{-2}\hat{x},
\eea
where $\hat{x}$ is the largest positive real root  of the equation
\bea
-d^6-2d^5 \sqrt{2 d+1} -4 d^5-2 d^4\sqrt{2 d+1} -2 d^4\nonumber \\
   +\left(4 d^4+5 d^3\sqrt{2 d+1} +10 d^3+5 d^2\sqrt{2 d+1}
   +5 d^2\right) x\nonumber \\
   +\left(-4 d^2-3 d\sqrt{2 d+1} -6d-3 \sqrt{2 d+1}-3\right) x^2+x^3=0
\eea

Using Eq.(\ref{ds0ps}) we obtain that the spectral dimension $d_S$ is therefore given by 
\bea
d_S=2\frac{\ln(d+1)}{\ln\lambda}.
\eea

\subsection{Case $m<d-3$}

In this paragraph we study the RG flow  for the   pseudo-fractal simplicial complex for  $m<d-3$.
By expressing Eqs. (\ref{RGps}) in terms of the variables $x_{\ell}^{(\tau)}$ defined in Eq. (\ref{xell}) and the variables $p_{\ell}^{(\tau)}$ calculated at iteration $\tau$, we obtain the recursive equations
\bea
&&\hspace*{-25mm}x_{\ell}^{(\tau+1)}=\left[x_{\ell+1}^{(\tau)}+\left(x_1^{(\tau)}-\frac{(m+1)}{m+(d-m) x_1^{(\tau)}}\right)\frac{[(d-m-2)^{\ell}-1]}{d-m-3}\right]\left[p_2^{(\tau)}+\frac{d-m-1}{m+(d-m)x_1^{(\tau)}}\right]^{-1},\nonumber \\
&&\hspace*{-25mm}p^{(\tau+1)}_{\ell}=\left[p_{\ell+1}^{(\tau)}+\frac{(d-m-1)}{m+(d-m)x_1^{(\tau)}}\frac{[(d-m-2)^{\ell}-1]}{d-m-3}\right]\left[p_2^{(\tau)}+\frac{d-m-1}{m+(d-m)x_1^{(\tau)}}\right]^{-1},
\label{RGhdps4}
\eea
with initial conditions  $x_{\ell}^{(0)}=1-\mu$ and $p_{\ell}^{(0)}=1$.
These equations can be solved in terms of the variables $y_1^{(\tau)}$ defined as 
\bea
{y}_{1}^{(\tau)}=p_{2}^{(\tau)}+\frac{(d-m-1)}{m+(d-m)x_1^{(\tau)}}.
\eea 
In particular the solution of Eqs. (\ref{RGhdps4}) is given by 
\bea
p_{2}^{(\tau+1)}&=&\prod_{m=1}^{\tau}\frac{1}{y_1^{(m)}}+\frac{(d-m-1)}{(d-m-3)}\sum_{m=1}^{\tau}\frac{[(d-m-2)^{\tau-m+2}-1]}{m+(d-m)x^{(m)}}\prod_{m'=m}^{\tau}\frac{1}{y_1^{(m')}}
,
\nonumber\\ 
y_1^{(\tau+1)}&=&p_2^{(\tau+1)}+\frac{(d-m-1)}{m+(d-m) x_1^{(\tau+1)}}\nonumber\\
&=&\prod_{m=1}^{\tau}\frac{1}{y_1^{(m)}}+\frac{(d-m-1)}{(d-m-3)}\sum_{m=1}^{\tau}\frac{[(d-m-2)^{\tau-m+2}-1]}{m+(d-m)x^{(m)}}\prod_{m'=m}^{\tau}\frac{1}{y_1^{(m')}}\nonumber \\
&&+\frac{(d-m-1)}{m+(d-m) x_1^{(\tau+1)}}
,
\nonumber \\
x_{1}^{(\tau+1)}&=&x_1^{(1)}\prod_{m=1}^{\tau}\frac{1}{y_1^{(m)}}+\frac{1}{d-m-3}\sum_{m=1}^{\tau}\left(x_1^{(m)}-\frac{m+1}{m+(d-m)x_1^{(m)}}\right)\nonumber \\
&&\times[(d-m-2)^{\tau+1-m}-1]\prod_{m'=m}^{\tau}\frac{1}{y_1^{(m')}}
.
\eea
In order to turn this system of equations into a Markovian system of equations, we again express the variables at iteration $\tau+1$ only in terms of variables at iteration $\tau$. We then have
\bea
y_1^{(\tau+1)}&=&A^{(\tau)}+B^{(\tau)}-D^{(\tau)}+\frac{(d-m-1)}{m+(d-m)x_1^{(\tau+1)}}
,
\nonumber\\
x_{1}^{(\tau+1)}&=&(1-\mu)A^{(\tau)}+C^{(\tau)}-E^{(\tau)}\nonumber 
\eea
with $A^{(\tau)},B^{(\tau)},C^{(\tau)},D^{(\tau)},E^{(\tau)}$ given by 
\bea
A^{(\tau)}&=&\prod_{m=1}^{\tau}\frac{1}{y_1^{(m)}}
,
\nonumber \\
B^{(\tau)}&=&=\frac{(d-m-1)}{(d-m-3)}\sum_{m=1}^{\tau}\frac{(d-m-2)^{\tau-m+2}}{m+(d-m)x^{(m)}}\prod_{m'=m}^{\tau}\frac{1}{y_1^{(m')}}
,
\nonumber \\
C^{(\tau)}&=&\frac{1}{d-m-3}\sum_{m=1}^{\tau}\left(x_1^{(m)}-\frac{(m+1)}{m+(d-m)x_1^{(m)}}\right)(d-m-2)^{\tau+1-m}\prod_{m'=m}^{\tau}\frac{1}{y_1^{(m')}}
,
\nonumber \\
D^{(\tau)}&=&\frac{(d-m-1)}{(d-m-3)}\sum_{m=1}^{\tau}\frac{1}{m+(d-m)x^{(m)}}\prod_{m'=m}^{\tau}\frac{1}{y_1^{(m')}}
,
\nonumber \\
E^{(\tau)}&=&\frac{1}{d-m-3}\sum_{m=1}^{\tau}\left(x_1^{(m)}-\frac{(m+1)}{m+(d-m)x_1^{(m)}}\right)\prod_{m'=m}^{\tau}\frac{1}{y_1^{(m')}}.
\label{defAdg3}
\eea
The RG flow can therefore be cast in a set of recursive equations for $A^{(\tau)},B^{(\tau)},C^{(\tau)},D^{(\tau)}$ and $E^{(\tau)}$ given by
\bea
y_1^{(\tau+1)}&=&A^{(\tau)}+B^{(\tau)}-D^{(\tau)}+\frac{d-m-1}{m+(d-m)[(1-\mu)A^{(\tau)}+C^{(\tau)}-E^{(\tau)}]}
,
\nonumber\\
x_{1}^{(\tau+1)}&=&(1-\mu)A^{(\tau)}+C^{(\tau)}-E^{(\tau)}
,
\nonumber \\
A^{(\tau+1)}&=&\frac{1}{y_1^{(\tau+1)}}A^{(\tau)}
,
\nonumber \\
B^{(\tau+1)}&=&\frac{d-m-2}{y_1^{(\tau+1)}}B^{(\tau)}+\frac{(d-m-1)(d-m-2)^2}{(d-m-3)}\frac{1}{y_1^{(\tau+1)}[m+(d-m))x_1^{(\tau+1)}]}
,
\nonumber \\
C^{(\tau+1)}&=&\frac{(d-m-2)}{y_1^{(\tau+1)}}C^{(\tau)}+\frac{(d-m-2)}{(d-m-3)}\frac{1}{y_1^{(\tau+1)}}\left(x_1^{(\tau+1)}-\frac{(m+1)}{m+(d-m)x_1^{(\tau+1)}}\right)
,
\nonumber \\
D^{(\tau+1)}&=&\frac{1}{y_1^{(\tau+1)}}D^{(\tau)}+\frac{(d-m-1)}{(d-m-3)}\frac{1}{y_1^{(\tau+1)}[m+(d-m)x_1^{(\tau+1)}]}
,
\nonumber \\
E^{(\tau+1)}&=&\frac{1}{y_1^{(\tau+1)}}E^{(\tau)}+\frac{1}{(d-m-3)}\frac{1}{y_1^{(\tau+1)}}\left(x_1^{(\tau+1)}-\frac{(m+1)}{m+(d-m)x_1^{(\tau+1)}}\right),
\eea
with initial conditions 
$A^{(1)},B^{(1)},C^{(1)},D^{(1)},E^{(1)}$ which can be found by inserting $x_1^{(0)}=1-\mu$ and $y_1^{(0)}=[1+\frac{d-m-1}{d-m\mu)}]$ in Eq. (\ref{defAdg3}).

By extracting the leading eigenvalue $\lambda$ close to the relevant fixed point at $\mu^{\star}=0$ and using Eq. (\ref{ds0ps}) we can deduce the values of the spectral dimension $d_S$ (see Table \ref{table:pseudofractal}).

Here we make an additional useful observation. As is true for the specific case  $m=0$ and $d>3$ (see Ref.\cite{RG1}) and in the more general case investigated here with $m<d-3$, we observe that the RG Eqs.(\ref{RGhdps4}) of the pseudo-fractal simplicial complex have the same leading term of the RG Eqs.(\ref{RGhd3}) valid for the Apollonian simplicial complex with $m<d-3$. Therefore the  leading eigenvalue $\lambda$ of the  Eqs.(\ref{RGhdps4}) is given by 
\bea
\lambda=\frac{ {d^2}-m(d+1)}{d^2-(m+1)(d+1)}+\mathcal{O}(d^{-1}).
\eea
It follows that for  $d\gg1 $ and $m$ finite, the spectral dimension $d_S$ obeys the asymptotic scaling 
\bea
d_S\simeq 2 (d-m) \log (d+1)+{\mathcal O}(\log(d)),
\eea
i.e. it grows faster than linearly with $d$.

\section{Main results  and comparison to numerical results}
\subsection{Higher-order spectral dimensions of Apollonian and pseudo-fractal simplicial complexes}

\begin{table}
\caption{\label{table:Apollonian}Numerical values for the spectral dimension $d_S$ of the $m$-up-Laplacian (with $m\leq d-3$) of the $d$-dimensional Apollonian simplicial complexes up to dimension $d=9$. The values of $d_S$ are rounded at the  sixth significant figure. 
}
\begin{indented}
\item[]\begin{tabular}{@{}lllllllll}
\br
$d$/$m$&$d=2$ &$d=3$& $d=4$& $d=5$ & $d=6$ & $d=7$ &$d=8$&$d=9$\\
\mr
$m=d-3$&-&3.73813&4.5742&5.19979&5.70072&6.11932&6.47949 &6.79596\\
$m=d-4$ &-&-&7.39962&8.48212&9.35664&10.0913&10.7253& 11.2833\\
$m=d-5$&-&-&-&11.729&12.9719&14.0179&14.9217&15.7178\\
$m=d-6$&-&-&-&-&16.5732&17.9293& 19.1017&20.1346\\
$m=d-7$&-&-&-&-&-&21.8337& 23.2741&24.5434\\
$m=d-8$&-&-&-&-&-&-&27.4423&28.9478\\
$m=d-9$&-&-&-&-&-&-&-&33.3496\\
\br
\end{tabular}
\end{indented}
\end{table}

\begin{figure}[h!]
	\begin{center}
 \includegraphics[width=1\columnwidth]{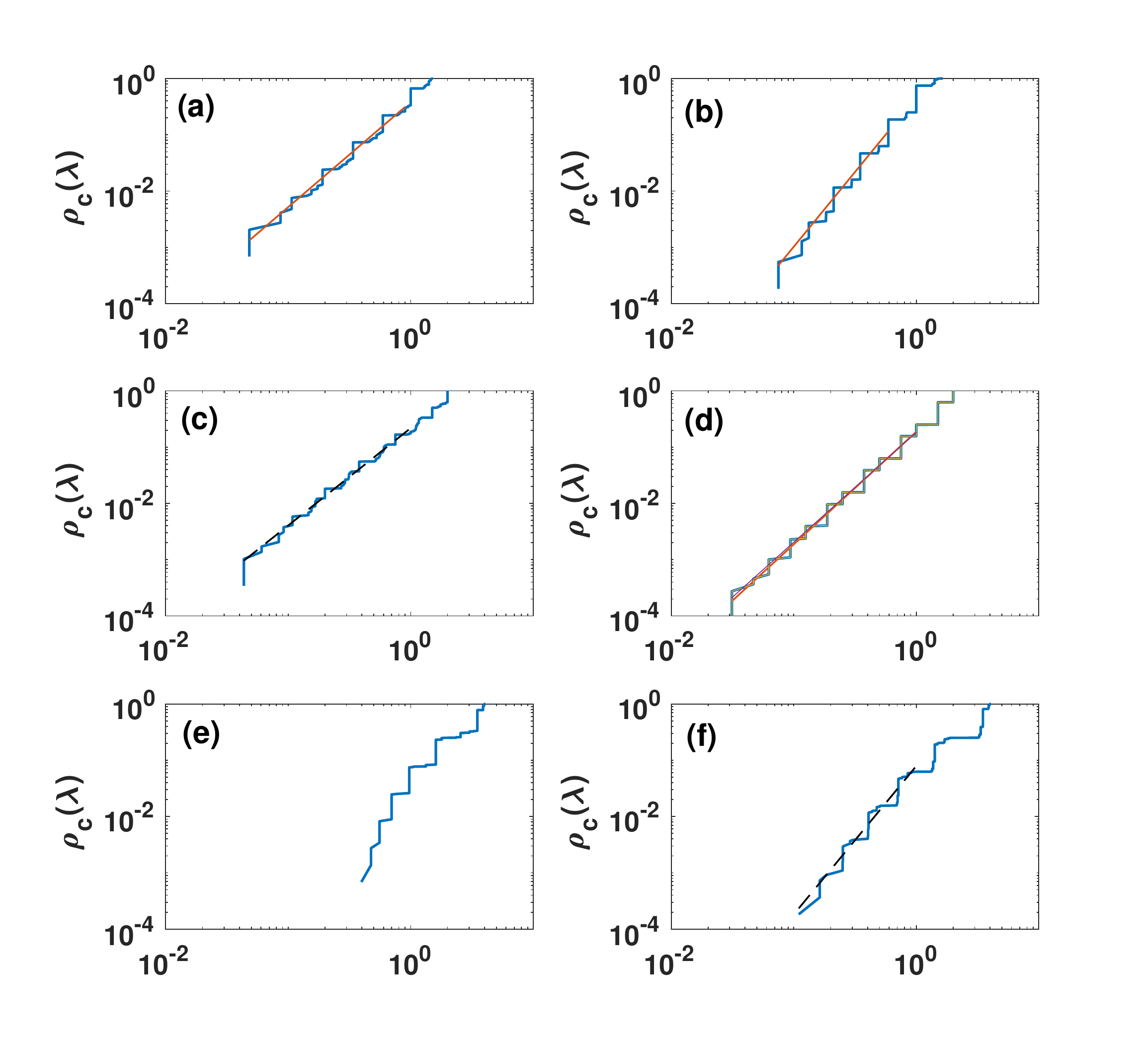}	
 \end{center}
 \caption{The cumulative density of eigenvalues $\rho_c(\lambda)$ of the up-Laplacians of order $m$ is shown in solid lines for the Apollonian and the pseudo-fractal simplicial complex of dimension $d=3$. Panels (a), (c) and (e) display results (blue lines) for the up-Laplacian of order $m$ of the  Apollonian simplicial complex with respectively $m=0,1,2$. Panels (b), (d) and (f) display results (blue lines) for the up-Laplacian of order $m$ of the  pseudo-fractal  simplicial complex with respectively $m=0,1,2$. The theoretically predicted spectral dimensions are shown with red lines. Dashed black lines indicate power-law fits. }
	\label{fig:Detd3}
	\end{figure}

\begin{figure}[h!]
	\begin{center}
 \includegraphics[width=1\columnwidth]{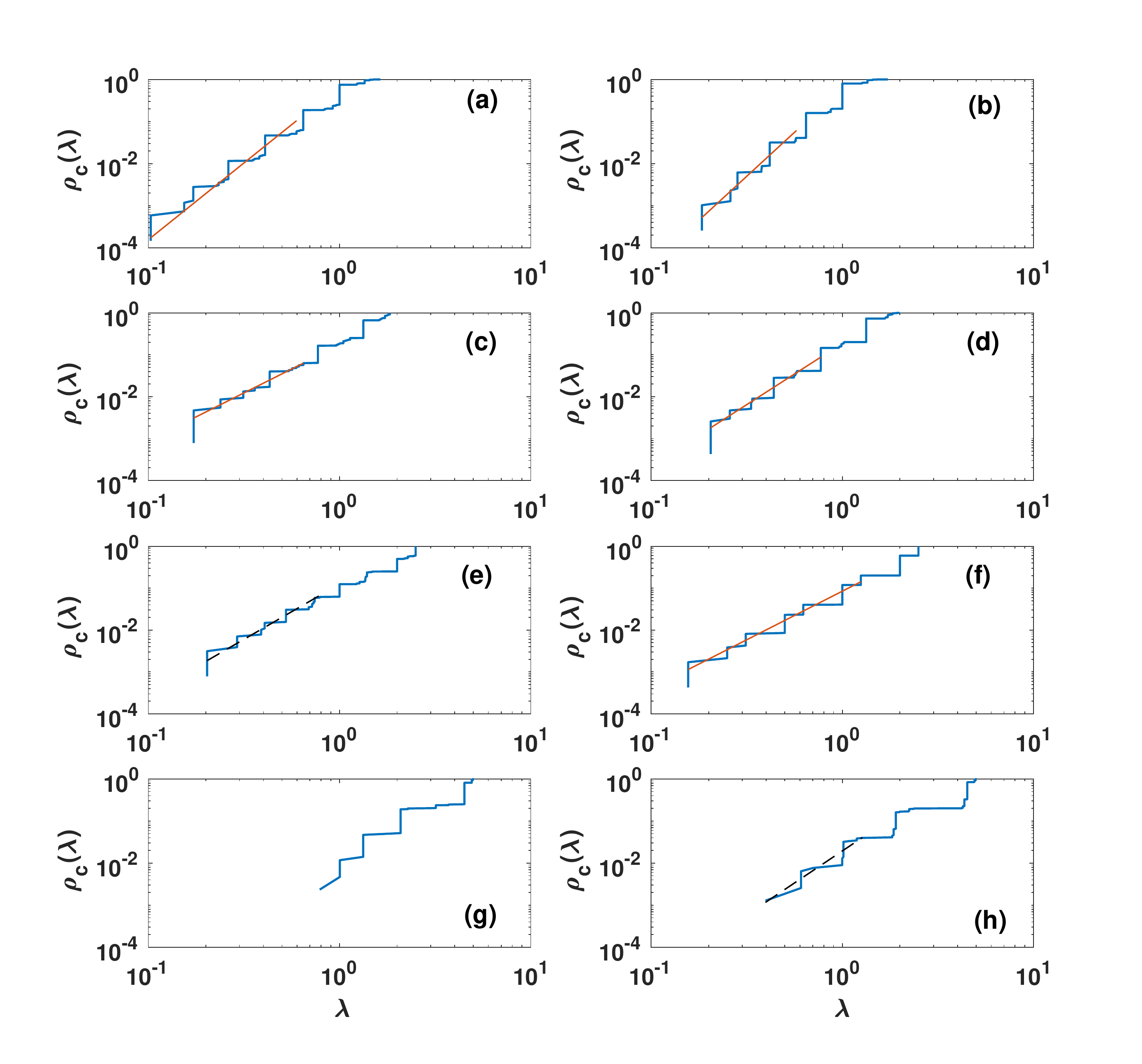}	
  \end{center}
  \caption{The cumulative density of eigenvalues $\rho_c(\lambda)$ of the up-Laplacians of order $m$ is shown in solid lines for the Apollonian and the pseudo-fractal simplicial complex of dimension $d=4$. Panels (a), (c), (e) and (g) display results (blue lines) for the up-Laplacian of order $m$ of the  Apollonian simplicial complex with respectively $m=0,1,2,3$. Panels (b), (d), (f) and (h) display results (blue lines) for the up-Laplacian of order $m$ of the  pseudo-fractal  simplicial complex with respectively $m=0,1,2,3$. The  theoretically predicted spectral dimensions are shown with red lines. Dashed black lines indicate power-law fits. }
	\label{fig:Detd4}
	\end{figure}

\begin{table}
\caption{\label{table:pseudofractal}
Numerical values for the spectral dimension $d_S$ of the $m$-up-Laplacian (with $m\leq d-2$) of the $d$-dimensional pseudo-fractal simplicial complexes up to dimension $d=9$. The values of $d_S$ are rounded at the  sixth significant figure. }
\begin{indented}
\item[]\begin{tabular}{@{}llllllllll}
\br
$d$/$m$&$d=2$ &$d=3$& $d=4$& $d=5$ & $d=6$ & $d=7$ &$d=8$&$d=9$\\
\mr
$m=d-2$&3.16993&4.0&4.64386& 5.16993 & 5.61471 &6.0 & 6.33985 &6.64386\\
$m=d-3$&-&5.31562&5.86924&6.28083&6.60535&6.87191&7.0975&7.29281\\
$m=d-4$ &-&-&8.37610&8.99732&9.49705&9.91547&10.276&10.5934\\
$m=d-5$&-&-&-&12.7140&13.7232&14.4689&15.057&15.5463\\
$m=d-6$&-&-&-&-&17.3048&18.5860&19.5562&20.3283\\
$m=d-7$&-&-&-&-&-&22.2618&23.7403&24.897\\
$m=d-8$&-&-&-&-&-&-&27.5667&29.1935\\
$m=d-9$&-&-&-&-&-&-&-&33.1841\\
\br
\end{tabular}
\end{indented}
\end{table}
In the preceding paragraphs we have derived the equations from which we can deduce the spectral dimensions $d_S$ of the up-Laplacians of order $m$ of the  Apollonian and pseudo-fractal simplicial complexes.  The only exceptions are the case $m=d-2$ for the Apollonian network and the case $m=d-1$ for the pseudo-fractal network.
The predicted values for the spectral dimensions $d_S$ for $d$-dimensional Apollonian ($m\leq d-3$) and pseudo-fractal simplicial complexes ($m\leq d-2$ ) up to dimension $d=9$ are shown in Table \ref{table:Apollonian} and Table \ref{table:pseudofractal} respectively.
In Figure \ref{fig:Detd3} and Figure \ref{fig:Detd4}) we compare the spectra obtained by numerical diagonalization of the higher-order up-Laplacians for Apollonian and pseudo-fractal simplicial complexes of dimension $d=3$ and $d=4$. We find a very good agreement with our exact analytical results.
In addition we can fit the numerical data finding the spectral dimensions for the case $m=d-1$ of the Apollonian simplicial complex and the case $m=d-2$ of the pseudo-fractal simplicial complex.

From our RG calculations of the spectrum of higher-order up-Laplacians of Apollonian simplicial complexes and pseudo-fractal simplicial complexes and its numerical validation we draw the following main conclusions:
\begin{itemize}
\item[(1)]
Higher-order up-Laplacians of order $m$ on Apollonian and pseudo-fractal simplicial complexes display a finite spectral dimension with the only exception of the case of $m=d-1$ for the Apollonian simplicial complex.
A single simplicial complex generated by the  {above}-mentioned models is therefore not just characterized by a single spectral dimension but by multiple spectral dimensions corresponding to different orders $m$.
\item[(2)]
The analytical prediction of the spectrum of the $m$-order up-Laplacian  on $d$-dimensional Apollonian and pseudo-fractal  simplicial complexes shows that the  spectral dimension $d_S$ decreases with increasing $m$ as long as $m\leq d-3$ for the Apollonian simplicial complexes and as long as $m\leq d-2$ for the pseudo-fractal simplicial complex.
\item[(3)]
The symmetries of the simplicial complex do not only induce   degenerate eigenvalues for the graph Laplacian \cite{RG1} but also for their higher-dimensional counterparts. 
Indeed, from our numerical results  (Figures \ref{fig:Detd3}) and \ref{fig:Detd4})) we observe  that the higher-order up-Laplacian have several eigenvalues that are highly degenerate. 
\end{itemize}

\section{Conclusions}

Higher-order Laplacians are important topological objects that generalize graph Laplacians and extend the notion of diffusion to higher dimension.
Here we show that two non-amenable simplicial complex models (the Apollonian simplicial complex, the pseudo-fractal simplicial complex) display finite higher-order spectral dimensions $d_S$.
We  observe  that  a single simplicial complex can be characterized by a set of spectral dimensions corresponding to the spectrum of the up-Laplacians of different order $m$.
We have used renormalization group methods applied to a Gaussian model to predict the higher-order spectral dimension $d_S$ of up-Laplacians of order $m$ of the Apollonian simplicial complex and  the pseudo-fractal simplicial complex of arbitrary dimension $d$. With our RG approach it is possible to analytically calculate the spectral dimension $d_S$ for order $m\leq d-3$ for the Apollonian simplicial complexes and for order $m\leq d-2$ for pseudo-fractal simplicial complexes.  In these cases the spectral dimensions are determined by the scaling of the RG flow  away from the repulsive fixed point at zero mass, i.e. at $(\mu_1^{\star},p_2^{\star})=(0,p^{\star})$.
Additionally we have found that in  the  range of values of $m$ for which we can predict the spectral dimension, the spectral dimension $d_S$ up-Laplacians of order $m$ decreases as $m$ increases. Our analytical calculations are validated by numerical results.
 {In the future  \cite{next} we plan to characterize the the higher-order Laplacians of the simplicial complex model called ``Network  Geometry with Flavor" in order to investigate the role  of randomness in determining the   spectral properties of the simplicial complexes and the  implications that topological phase transitions  have on higher-order spectra.}
We hope that the present work can stimulate further research on higher-order spectral dimensions and topological phase transitions in different fields related to network topology including quantum gravity and brain research.
\section*{Acknowledgements}

This research was supported in part by Perimeter Institute for Theoretical Physics. Research at Perimeter Institute is supported by the Government of Canada through the Department of Innovation, Science, and Economic Development, and by the Province of Ontario through the Ministry of Research and Innovation. M. R. was partly supported through a Projectruimte grant of the Netherlands Organisation for Scientific Research (NWO).

\end{document}